\documentclass[aps,prd,onecolumn,superscriptaddress]{revtex4}  % for review and submission 
\usepackage{graphicx}  % needed for figures
\usepackage{dcolumn}   % needed for some tables
\usepackage{bm}        % for math
\usepackage{amssymb}   % for math
\usepackage{amsmath}   % for math
\usepackage{multirow}
\usepackage{color}
\usepackage{units}

\usepackage{amssymb}
\usepackage{amsmath}

\newcommand{\cm}{{\cal M}}

\begin{document}

% Title
\title{Status of Searches for Magnetic Monopoles}
\thanks{This review is dedicated to the memory of Giorgio Giacomelli, who devoted much of his
scientific activity to the study of magnetic monopoles. }

%Authors, affiliations address.
\author{L. Patrizii$^1$ and M. Spurio}
\affiliation{Istitito Nazionale di Fisica Nucleare - Sezione di Bologna -
Viale Berti Pichat 6/2 - 40127 Bologna (Italy)}
\affiliation{Dipartimento di Fisica e Astronomia dell'Universit\`a di Bologna - Viale Berti Pichat 6/2 - 40127 Bologna (Italy)}  

%Abstract
\begin{abstract}
The searches for magnetic monopoles ($\cm$s) is a fascinating interdisciplinary field with implications in fundamental theories, in particle physics,  astrophysics, and  cosmology. 
The quantum theory of $\cm$s and its consistency with electrodynamics was derived by P.A.M. Dirac.
This marked the start of the searches for \textit{classical monopoles} at every new accelerator, up to the LHC.
Magnetic monopoles are required by Grand Unification Theories, but unlike  classical monopoles they would be incredibly massive, out of the reach of any conceivable accelerator.
Large efforts have been made to search for them in the cosmic radiation as relic particles from the early Universe in the widest range of mass and velocity experimentally accessible.

In this paper the status of the searches for classical $\cm$s at accelerators, for GUT, superheavy $\cm$s in the penetrating cosmic radiation and for Intermediate Mass $\cm$s at high altitudes is discussed, with emphasis on the most recent results and future perspectives.

\end{abstract}

%Keywords, etc.
%\begin{keywords}
%magnetic monopoles; dyons; proton decay catalysis; accelerator searches; cosmic ray searches 
%\end{keywords}
\maketitle

%Table of Contents
%%%\tableofcontents

%%%%%%%%%%%%%%%%%%%%%%%%%%%%%%%%%%%%%%%%%%%%%%%%%%%%%%%%%%%%%%%%%%%%%%
\section{Introduction}
%{\it\small(2 pages)}
\label{sec:Introduction}
%% {}``gas '' 
%%%%%%%%%%%%%%%%%%%%%%%%%%%%%%%%%%%%%%%%%%%%%%%%%%%%%%%%%%%%%%%%%%%%%%

%%% MM: Classici
%%%%%%%%%%%%%%%%%%
% Margin Note
%\begin{marginnote}[120pt]
%\entry{Classical Magnetic Monopoles}{ }
%\end{marginnote}
The concept of the magnetic monopole ($\cm$) was introduced by P.~A.~M. Dirac in 1931 \cite{dirac}  to explain the quantization of the electric charge. 
Dirac established the basic relation between the elementary electric charge $e$ and the magnetic charge $g$
\begin{equation}\label{eq:1.g}
{eg\over c}={n\hbar \over 2 } \longrightarrow g = n\cdot g_D = n\cdot {1\over 2} {\hbar c \over e} \sim n\cdot {137\over 2} e \ ,
\end{equation}
where $g_D$ is the unit Dirac charge and $n$ is an integer.
The existence of magnetic charges and currents would symmetrize the Maxwell's equations. 
The symmetry would not be perfect, since the unit magnetic charge is much larger than the electric one. 
Eq. \ref{eq:1.g} defines most of the properties of Dirac (or classical) magnetic monopoles, as they are assumed as point-like particles. 

%%% MM: GUT
%%%%%%%%%%%%%%%%%%
Around 1974 it was realized \cite{74H1, 74P2} that the electric charge is naturally quantized in Grand Unification Theories (GUTs) of the strong and electroweak interactions.
$\cm$s appear at the phase transition corresponding to the spontaneous breaking of the unified group into subgroups, one of which is $U(1)$, which describes electromagnetism.
In a certain sense, the situation was reversed compared to the reasoning of Dirac: the quantization of the electric charge now implied the existence of magnetic monopoles.

%%% GUT
%%%%%%%%%%%%%%%%%%
In the contest of GUT's, the lowest $\cm$ mass, $M$, is related to the mass of the $X$ vector boson ($m_X\sim 10^{15}$ GeV), which is the carrier of the unified interaction and defines the unification scale,
$M \gtrsim m_X/\alpha \simeq 10^{17}$ GeV, where $\alpha\simeq 0.025$ is the dimensionless unified coupling constant (natural units with $c=1$ in the c.g.s. system are almost always used in this paper).
Larger $\cm$ masses are expected if gravity is brought into the unification picture, and in some SuperSymmetric models.
%%% $\cm$: IMM
%%%%%%%%%%%%%%%%%%
Magnetic monopoles with masses $M\sim 10^{5} \div 10^{12}$ GeV (also called \textit{Intermediate Mass Monopoles}, IM$\cm$s) are predicted by theories with an intermediate energy scale between the GUT and the electroweak (EW) scales and would appear in the early Universe at a considerably later time than the GUT epoch \cite{laza-imm}.
The lowest mass $\cm$ is stable since magnetic charge is conserved, like electric charge. 

%%% Collider
%%%%%%%%%%%%%%%%%%
The possibility that the gauge theory of EW interactions allow magnetic monopole solutions (also with multiply magnetic charge) was realized more recently \cite{cho97,yang98}. 
With a basic magnetic charge twice the Dirac charge ($g=2g_D$), such $\cm$s can be interpreted as a non-trivial hybrid between the Dirac and the GUT (electrically charged) monopoles. 
The mass of this $\cm$ (dyon) has been estimated to range between 3 to 7 TeV \cite{cho12} which makes it a very good candidate for searches at the CERN Large Hadron Collider (LHC).

%%%%%%%%%%%%%%%%%%%%%

%%% In questo paper
%%%%%%%%%%%%%%%%%%
In this paper, the status and perspectives of the searches for magnetic monopoles are reviewed. 
The theoretical motivations for Dirac and GUT magnetic monopoles have been extensively covered in \cite{pres84} and only short recalls are considered here, \S \ref{sec:gauge}. 
An exhaustive review of the phenomenology of $\cm$s and of the early searches is in \cite{gg84}.
More recent reviews on the different monopole models are in \cite{Milton,raja}.
The interactions of $\cm$s with matter and their energy loss mechanisms are discussed in \S \ref{sec:3}, and the most common experimental techniques for their detection are presented in \S \ref{sec:4-tech}.
The status of the searches for $\cm$s in the cosmic radiation and at accelerators is discussed in \S \ref{sec:searches}.
Finally, the perspectives for on-going and future experiments are in \S \ref{sec:perspective}. 

%%%%%%%%%%%%%%%%%%%%%%%%%%%%%%%%%%%%%%%%%%%%%%%%%%%%%
%%%%%%%%%%%%%%%%%%%%%%%%%%%%%%%%%%%%%%%%%%%%%%%%%%%%%
%\section{Properties of Magnetic Monopoles}
\section{Classical, GUT and Intermediate Mass Magnetic Monopoles}
%{\it\small( 4 pages)}
\label{sec:gauge}
%%{}``gas ''
%%%%%%%%%%%%%%%%%%%%%%%%%%%%%%%%%%%%%%%%%%%%%%%%%%%
%%%%%%%%%%%%%%%%%%%%%%%%%%%%%%%%%%%%%%%%%%%%%%%%%%%%%

%\begin{marginnote}[1pt]
%\entry{1}{pippo}
%\end{marginnote}

No predictions for the Dirac monopole mass, $M_D$, exist. 
A kind of rule of thumb assumed that the classical electron radius be equal to the {}``classical monopole radius'', from which one has $M_D\sim g_D^2 m_e/e^2 \sim 4700 m_e \sim 2.4$ GeV.
Based on this assumption, classical $\cm$s were searched for at every new accelerator. 

At accelerators $\cm$s could be produced via electromagnetic processes similar to those involving the creation of electrically charged fermion pairs, through a dimensionless magnetic constant $\alpha_m$ dependent on the magnetic unit charge $g_D$. 
This constant may be introduced in analogy with the fine-structure constant $\alpha= e^2/\hbar c\simeq 1/137$ and corresponds to $\alpha_m =g_D^2/\hbar c \simeq 34.25$. Therefore, the electromagnetic interactions of $\cm$s are too strong for perturbative theory to be applicable.
A quantitative estimate of the production cross sections cannot be reliably derived and it remains an open problem.

GUTs and IM$\cm$s are expected to be very massive composite objects, well beyond the reach of any existing or foreseen accelerator. 
Such poles could have been produced in the early Universe \cite{kibble}, be a component of the cold dark matter, and still exist as cosmic relics. 
Their kinetic energy would have been affected by the Universe expansion and by travelling through the galactic and intergalactic magnetic fields.

%%%%%%%%%%%%%%%%%%%%%%%%%%%%%%%%%%%%%%%%%%%%%%%%%%%%%%%%%%%
\subsection{History of GUT Magnetic Monopoles}
\label{sec:GUTMM}
%%%%%%%%%%%%%%%%%%%%%%%%%%%%%%%%%%%%%%%%%%%%%%%%%%%%%%%%%%%
%% GG+LP arxives
%%%%%%%%%%%%%%%%%% 

In the standard Big Bang model, the Universe started in a state of extremely large density and temperature. 
As time progressed, density and temperature decreased and the particle composition changed.
The grand unification of strong and electroweak interactions lasted until the temperature dropped to $\sim 10^{15}$ GeV, when a phase transition is thought to have occurred. 
During that phase transition, GUT monopoles appeared as topological defects, about one pole for each causal domain. 
This would lead to a present-day $\cm$ abundance \cite{pre79}, exceeding by many orders of magnitude the critical energy density of the Universe \cite{pdg}:
\begin{equation}\label{eq:2.rhoc}
\rho _{c}\equiv\frac{3H^{2}}{8\pi G_{N}}
= 1.88\cdot 10^{-29} { h}^{2}\ \textrm{ [g cm}^{-3}]
= 1.05\cdot 10^{-5} { h}^{2}\ \textrm{ [GeV cm}^{-3}] \ ,
\end{equation}
the so-called \textit{magnetic monopole problem}.
In Eq. \ref{eq:2.rhoc} the scaled Hubble parameter, $h\sim 0.7$, is defined in terms of the Hubble constant $H\equiv 100 h$ km s$^{-1}$ Mpc$^{-1}$. 
Cosmological inflation would lead to present-day $\cm$'s abundance significantly smaller than could be plausibly detected. 
The reduction of $\cm$s in the Universe was one of the motivating factors for cosmological inflation in Guth's original work \cite{guth}.
Observable abundances can be obtained in scenarios with carefully tuned parameters or if the reheating temperature was large enough to allow $\cm$'s production in high-energy collisions of fermions $f$, like $f \overline f \rightarrow \cm \overline \cm$.

%% RNCGG
As the Universe expanded and cooled down the energy of $\cm$s decreased for the adiabatic energy loss, like any other particle and radiation.
$\cm$s would have reached a speed $\beta \sim 10^{-10}$ during the epoch of galaxy formation.
As matter (including $\cm$s and dark matter) started to condense gravitationally into galaxies, galactic magnetic fields developed through the dynamo mechanism. 
$\cm$s were accelerated by these magnetic fields and after times of the order of $10^7$ y they could be ejected with velocities $\beta \sim (1\div 3)\times 10^{-2}$.
This process would give rise to an isotropic intergalactic flux of relatively high-energy $\cm$s.

Today we may expect to have a sizable fraction of monopoles bound to different \textit{systems}, as the galaxy cluster, our Galaxy or the Solar system.
In the absence of magnetic fields, $\cm$s would have velocities comparable to the virial velocity of objects gravitationally bound to the \textit{system}. 
There may be peaks in the velocity spectrum corresponding to poles bound to the Sun ($\beta  \sim 10^{-4}$), to the Galaxy ($\beta  \sim 10^{-3}$), or to the local galaxy cluster ($\beta  \gtrsim 10^{-2}$).
These gravitationally bound $\cm$s could produce a flux on Earth.

A magnetic field $\mathbf{B}$ acting over a length $\ell$ increases the kinetic energy of a monopole by a quantity $gB \ell$. 
The speed a $\cm$ can reach in a typical galactic magnetic field of strength $B \sim 3\ 10^{-6}$ G and a coherent length $\ell \sim 300 \textrm{ pc} = 10^{21}$ cm depends on its mass $M$, namely: \begin{equation}\label{Eq:2.v}
\beta \simeq \left\{
\begin{array}{ll}
c & M \lesssim 10^{11} \textrm{ GeV} \\
10^{-3} \big( { 10^{17}\textrm{ GeV}\over M }\big)^{1/2} & M \gtrsim 10^{11} \textrm{ GeV}
\end{array}\right.
\end{equation}
From the experimental point of view, classical cosmic monopoles are expected to be always relativistic, while GUT monopoles can have speeds over a wide range.

%\subsubsection{Intermediate Mass MMs}
%% 1410.1374%%%%%%%%%%%%%%%% 
Intermediate Mass Monopoles (IM$\cm$) also would survive as relics of the early Universe \cite{keph01}, and be stable.  
Because IM$\cm$s would be produced after inflation, their number would not be reduced by inflationary mechanisms.
IM$\cm$s with $10^{7} < M < 10^{12}$ GeV could acquire nearly-relativistic velocities in one coherent domain of the galactic field \cite{bhatta}.
Experimentally one has to look for IM$\cm$s with $\beta \gtrsim 0.1$.

%%%%%%%%%%%%%%%%%%%%% 
\subsubsection{Structure of a GUT Magnetic Monopole}
%% 1410.1374
%%%%%%%%%%%%%%%%%%%%% 

%%%%%%%%%%%%%%%%%%%%%%%%%%%%%%%%%%%%%%%%%%%%%%%%%%%%%%%%%%%
\begin{figure*}[tb]
%\vspace*{2mm}
\begin{center}
\includegraphics[width=14.0cm]{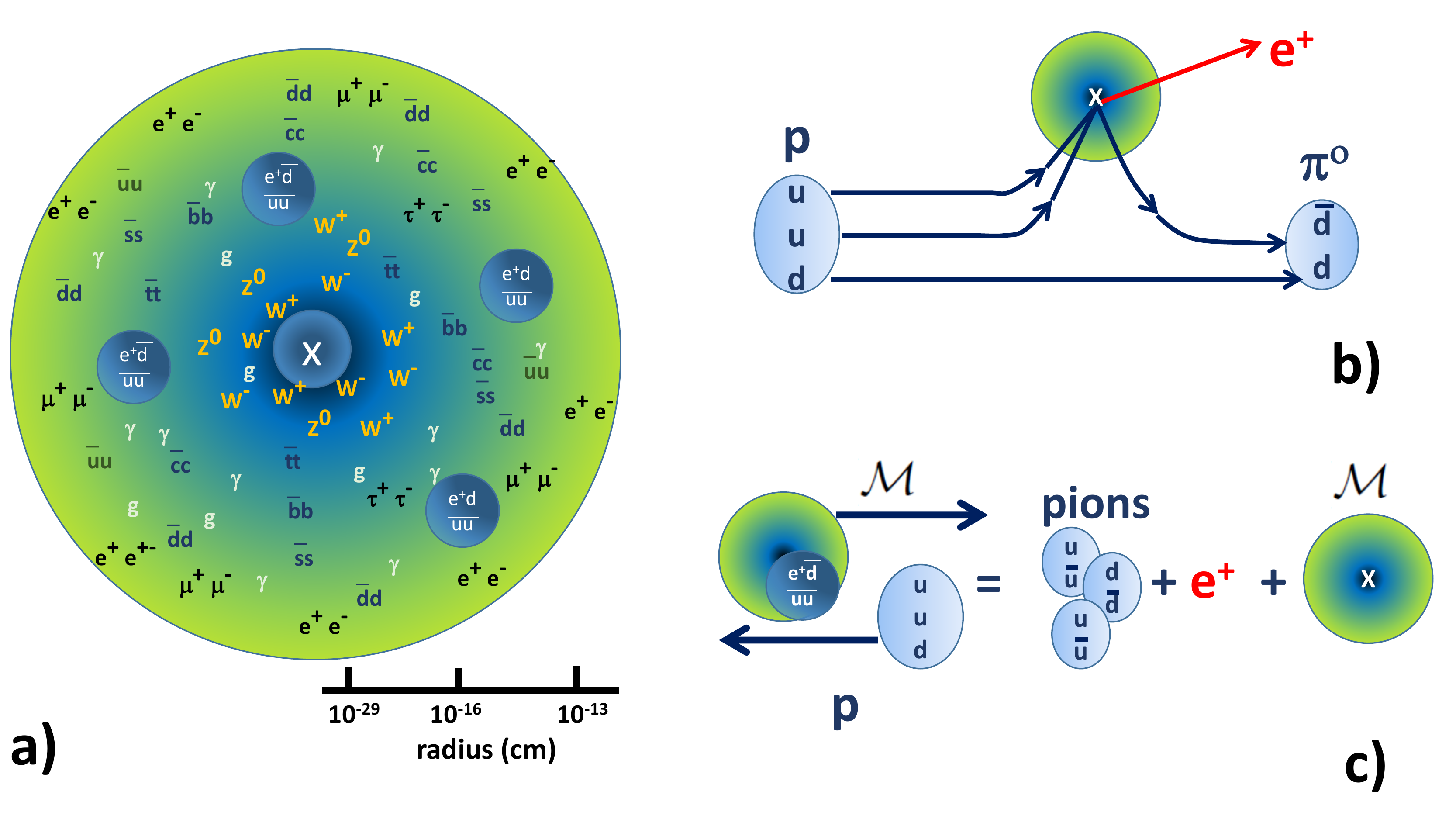}
\end{center}
\caption{\small\label{fig:inner} a) Illustration of the GUT monopole structure. The sketch illustrates various regions corresponding to 
$i)$ grand unification ($r\sim 10^{-29}$ cm): inside this core virtual $X$-bosons should be present; 
$ii)$ electroweak unification ($r\sim 10^{-16}$ cm) with virtual $W^\pm,\  Z^0$-bosons; 
$iii)$ the confinement region ($r\sim 10^{-13}$ cm) with virtual photons, gluons and a condensate of fermion-antifermion, four-fermion bags. 
For radii larger than few fm one has the field of a point magnetic charge $B = g/r^2$.
b) Illustrations of the monopole catalysis of proton decay through the reaction $ p + \cm \rightarrow \cm + e^+ +\pi^0$. c) The effect of the presence of a condensate of 4-fermions $\overline u \overline u \overline d e^+$ able to induce the proton decay is shown.}
\end{figure*}
%%%%%%%%%%%%%%%%%%%%%%%%%

The central core of a GUT monopole retains the original symmetry and contains the fields of the superheavy gauge bosons, which mediate baryon number violation. 
Through these bosons the strong and EW interactions between particles are indistinguishable: quarks and leptons are, in this domain, the same particles. 
The $\cm$'s core should be surrounded by a fermion-antifermion condensate. The condensate would have baryon number violating terms (see \S \ref{sec:2cata}) extending up to the confinement region. 

Fig. \ref{fig:inner} shows the structure of a GUT $\cm$: a very small core, an electroweak region, a confinement region, a fermion-antifermion condensate (which may contain four-fermion agglomerates that induce baryon number violating processes).
For $r \ge 3$ fm a GUT $\cm$ behaves as a point particle generating a field $B = g/r^2$.

IM$\cm$s produced after the decoupling of the strong and EW interactions would have a structure similar to that of a GUT $\cm$.
The core region would be larger (since $R \sim 1/M$) and the outer cloud would not contain terms that allow baryon number violation.

%%%%%%%%%%%%%%%%%%%%% 
\subsubsection{Dyons}
%% and $\cm$-p bound states gg
%%%%%%%%%%%%%%%%%%%%% 
A dyon is a particle carrying both electric and magnetic charges. Therefore, when at rest, it produces both electrostatic and magnetostatic fields. 
Considering two dyons, one with charges $e_1, g_1$ and the other with $e_2, g_2$, the Dirac quantization condition becomes
$e_1g_2-e_2g_1 = ({\hbar c / 2}) n $.
Semi-classical arguments were used to argue that in a proper quantum-mechanical treatment the dyon charge must be quantized. 
In the GUT framework quantum fluctuations lead to a quantized electric charge of the dyon, in integer multiples of the minimal electric charge \cite{dyons}.

%%% APP_Scin
A magnetic monopole and a proton may form a bound system with both magnetic and electric charges \cite{83B1}. As far as the energy loss in matter is concerned, a positively charged dyon and a $(\cm + p)$ bound system behave in the same way.

%%%%%%%%%%%%%%%%%%%%% 
\subsubsection{Magnetic monopole catalysis of nucleon decay\label{sec:2cata}}
%% 1410.1374
%%%%%%%%%%%%%%%%%%%%% 
As early as 1980 it was hypothesized that, given its inner structure, a GUT $\cm$ could catalyze baryon number violating processes, such as $ p + \cm \rightarrow \cm + e^+ + \textrm{ mesons}$.
The cross section $\sigma_0$ of this process would be incredibly small if of the order of the geometrical dimension of the $\cm$ core ($\sim 10^{-58}$ cm$^2$).
If the interaction is independent of $m_X$ the corresponding cross section $\sigma_0$ could be comparable to that of ordinary strong interactions, the so-called Rubakov-Callan mechanisms \cite{81R1,82C4}.
This possibility has important implications both in particle physics and in astrophysics.
The catalysis reaction on the $\cm$ core can be imagined pictorially as shown in Fig. \ref{fig:inner} b), while the effect of the presence of a condensate of 4-fermions $\overline u \overline u \overline d e^+$ is illustrated in Fig. \ref{fig:inner} c).

The cross section for the catalysis of nucleon decay, $\sigma_{cat}$, depends not only on the cross section $\sigma_0$, but also on the monopole speed as $\sigma_{cat}= (\sigma_0 /\beta) \cdot F(\beta)$.
The correction factor $F(\beta)$ takes into account an additional angular momentum of the monopole-nucleus-system and becomes relevant for speeds below a threshold $\beta_0$, which depends on the nucleus \cite{cata_b}. 
For speeds above $\beta_0$, $F(\beta)=1$. 

Positively charged dyons, or ($\cm+p$) states, should not catalyze proton decay with large rates because of the electrostatic repulsion between the proton and the dyon. Negatively charged dyons would instead have large effective cross sections.

%%%%%%%%%%%%%%%%%%%%%%%%%%%%%%%%%%%%%%%%%%%%%%%
\subsection{Cosmological and astrophysical bounds on GUT monopoles}
%% RNCgg84
%%%%%%%%%%%%%%%%%%%%%%%%%%%%%%%%%%%%%%%%%%%%%%%
Different upper limits on the flux of cosmic $\cm$s have been obtained based on cosmological and astrophysical considerations.
Most of these bounds have to be considered as rough orders of magnitude only.
A simple cosmological bound is obtained by requiring that the present monopole mass density, $\rho_M $, be smaller than the critical density given in Eq. \ref{eq:2.rhoc}, $\rho_M = n_M M <\rho_c$.
Correspondingly one should have for the $\cm$ number density $n_M <\rho_c/M$.
The $\cm$ flux per unit solid angle, $\Phi_M$, is simply related to the number density $n_M$ and to the $\cm$s average speed $v=c\beta$ in the observed frame:
\begin{equation}\label{eq:2.phi}
\Phi_M= {n_M v \over 4 \pi} < {\rho_c c \over 4 \pi} \cdot {\beta \over M} \simeq 2.5\cdot 10^{-13} { h}^{2} \cdot {\beta \over (M/10^{17} \textrm{ GeV}) }
 \textrm{ [cm}^{-2} \textrm{s}^{-1} \textrm{sr}^{-1}] \ .
\end{equation}
The limit holds in the hypothesis of poles uniformly distributed in the Universe.

%\subsubsection{Limits from galactic and intergalactic magnetic fields}
%%%%%%%%%%%%%%%%% 
A more stringent limit for most of the $(M,\beta)$ parameter space can be derived by astrophysical considerations.
Most celestial bodies possess large-scale magnetic fields. 
In our Galaxy, the magnetic field is stretched in the azimuthal direction along the spiral arms, and it is likely originated by the non-uniform rotation of the Galaxy. 
The time scale of this generation mechanism is approximately equal to the rotation period of the Galaxy, $\tau_B \sim 10^8$ y. 
Magnetic poles would gain energy, which is taken from the stored magnetic energy. 
An upper bound for the $\cm$ flux can be obtained by requiring that the kinetic energy gained per unit time is at most equal to the magnetic energy generated by the dynamo effect. 
The rate of energy, $dE_M/dt$, acquired per unit volume $V$ by $\cm$s in a magnetic field $\mathbf{B}$ is
\begin{equation}\label{eq:2.din1}
{dE_M\over dt dV} = \mathbf{J}_M\mathbf{\cdot B} 
\quad \textrm{[erg s}^{-1} \textrm{cm}^{-3}] \ ,
\end{equation}
where the magnetic current density is $ \mathbf{J}_M = g n_M \mathbf{v}$ and $\mathbf{v}$  is the average pole velocity. 
The magnetic energy density generated by the dynamo effect per unit time gives
\begin{equation}\label{eq:2.din2}
{dE_D\over dt dV} = {\rho_B \over \tau_B} = {B^2 \over 8\pi \tau_B} \quad \textrm{[erg s}^{-1} \textrm{cm}^{-3}] \ ,
\end{equation}
where $\rho_B = B^2/8\pi$ is the energy density of the magnetic field $\mathbf{B}$.
Assuming $\mathbf{v}$  and $\mathbf{B}$ parallel over large distances,  the condition $ \mathbf{J}_M \cdot  \mathbf{B} < {B^2 /8\pi \tau_B}$ leads to the upper bound: $n_M <{B / (8\pi \tau_B g v)}$. 
As in Eq. \ref{eq:2.phi}, a relation between flux and number density can be obtained
\begin{equation}\label{eq:2.din4}
\Phi_M = {n_M v \over 4 \pi} \lesssim {B \over 32 \pi^2 \tau_B g} \simeq 10^{-16} \ \textrm{ [cm}^{-2} \textrm{s}^{-1} \textrm{sr}^{-1}]
\end{equation}
for the typical value of the Galactic magnetic field strength, $B \sim 3\ 10^{-6}$ G, and single Dirac charge, $g=g_D$. 
This is the so-called \textit{Parker bound} \cite{parker}.
Note that the condition in Eq. \ref{eq:2.din4} is always more stringent than Eq. \ref{eq:2.phi} for $M<10^{17}$ GeV and $\beta>10^{-3}$.

In a more detailed treatment \cite{82T1}, which assumes reasonable choices for the astrophysical parameters and random relation between $\mathbf{v}$ and $\mathbf{B}$, the $\cm$s would acquire smaller energies. Correspondingly, less energy is removed from the galactic field and the Parker bound is less restrictive. 
In different speed regimes, indicated in Eq. \ref{Eq:2.v}, and defining a critical speed $\beta_c=10^{-3}$,  the upper bounds derived in \cite{82T1} are
\begin{equation}\label{eq:2.din6}
\Phi_M  \lesssim \left\{
\begin{array}{ll}
10^{-15} \quad\quad\quad\quad\quad\quad\quad  \textrm{ [cm}^{-2} \textrm{s}^{-1} \textrm{sr}^{-1}],\quad & M \lesssim 10^{17} \textrm{ GeV} \\
%% & \\
10^{-15} \big( { 10^{17}\textrm{ GeV}\over M }\big) \big( {\beta \over\beta_c} \big)\ \textrm{ [cm}^{-2} \textrm{s}^{-1} \textrm{sr}^{-1}],\quad  & M \gtrsim 10^{17} \textrm{ GeV}
\end{array}\right.
\end{equation}
An \textit{extended Parker bound} obtained by considering the survival of an early seed field \cite{adams93,lewis00}, yields a tighter bound
\begin{equation}\label{eq:2.din7}
\Phi_M \lesssim 1.2\ 10^{-16} \biggl( {M \over 10^{17} \textrm{ GeV} }\biggr) \ \
\textrm{ [cm}^{-2} \textrm{s}^{-1} \textrm{sr}^{-1}]  \ .
\end{equation}

%%%%%%%%%%%%%%%%%%%%%%%%%%%%%%%%%%%%%%%%%%%%%%%%%%%%%%%%%%%
%%%%%%%%%%%%%%%%%%%%%%%%%%%%%%%%%%%%%%%%%%%%%%%%%%%%%%%%%%%
%%%%%%%%%%%%%%%%%%%%%%%%%%%%%%%%%%%%%%%%%%%%%%%%%%%%%%%%%%%
\section{Magnetic Monopole Energy Losses}\label{sec:3}
%{\it\small( 6 pages)}
%%{}``gas ''
%%%%%%%%%%%%%%%%%%%%%%%%%%%%%%%%%%%%%%%%%%%%%%%%%%%%%%%%%%%
%%%%%%%%%%%%%%%%%%%%%%%%%%%%%%%%%%%%%%%%%%%%%%%%%%%%%%%%%%%
%%%%%%%%%%%%%%%%%%%%%%%%%%%%%%%%%%%%%%%%%%%%%%%%%%%%%%%%%%%

The study of the $\cm$'s interactions is important to understand the stopping power mechanisms in matter in general and the energy losses in particle detectors in particular. 
The long-range interaction of a $\cm$ with a fermion, an atomic system or a nucleus is due to the ``magnetostatic'' interaction between the field, $\mathbf{B_M}$, due to the pole magnetic charge and the magnetic-dipole moment $\boldsymbol\mu$ of the fermion, atom or nucleus \cite{77K1, 83B1}.
The interaction energy arising from the magnetic charge--magnetic dipole interaction is given by
\begin{equation}\label{eq:3.i}
W_D = - \boldsymbol\mu \mathbf{\cdot B_M} \ ,
\end{equation}
which, for an electron, corresponds to $W_D= \hbar^2/4 m_e r^2$. Assuming a relative distance equal to the Bohr radius, $r= a_o= 0.53 \times 10^{-8}$ cm, $W_D\simeq 7$ eV, comparable to the binding energy of an atom.
Thus one expects a sizable deformation of an atom when a $\cm$ passes inside or close to an atomic system.
For a proton at a distance of $r=1$ fm from the monopole, $W_D \sim 30$ MeV, a value larger than the binding energy of nucleons in nuclei. Thus, one expects deformations of the nucleus when a monopole passes close to it.

Classical and ``light'' ($M<10^{11}$ GeV) cosmic $\cm$s are expected (Eq. \ref{Eq:2.v}) with relativistic velocities and huge energy losses, more than enough to have them easily detected.
Furthermore, their energy loss would be sufficiently large to stop a considerable fraction of monopoles crossing the Earth. 
Therefore, searches for monopoles trapped in the Earth matter are of particular importance. 
GUT $\cm$s have such large masses that it is difficult to accelerate them to large velocities and the study of the energy loss of slow moving $\cm$s is relevant for their detection.

%%%%%%%%%%%%%%%% 
\subsection{Stopping power of fast monopoles with $\beta>0.05$ }
\label{sec:3-eloss-1}
%%%%%%%%%%%%%%%% 
A moving magnetic monopole produces an electric field whose lines of force lie in a plane perpendicular to the monopole trajectory. 
In matter, this field may ionize or excite the nearby atoms or molecules.
The interaction of magnetic monopoles having velocities $\beta> 0.05$ and $\gamma \le 100$ with the electrons of a material is well understood. 
The ionization energy loss for electric charges, see \S \textit{Passage of particles through matter} of \cite{pdg}, can be adapted for a magnetic charge \cite{ahlen} by replacing $Ze \rightarrow g \beta$, and the stopping power becomes
\begin{equation}\label{eq:3.elos}
{dE\over dx} = {4\pi N_e g^2 e^2 \over m_e c^2} \biggl[ \ln \biggr( {2m_ec^2 \beta^2\gamma^2 \over I}\biggr) - {1\over 2} +{k \over 2} - {\delta \over 2} - B_m \biggr] \ ,
\end{equation}
where $N_e$ is the density of electrons, $m_e$ the electron mass, $I$ the mean ionization potential of the crossed medium, $\delta$ the density effect correction, $k$ the QED correction, and $B_m$ the Bloch correction (refer to \cite{derk98} for the numerical values for different materials).
Practically, in this velocity range a $\cm$ with magnetic charge $g_D$ behaves as an equivalent electric charge $(Ze)^2 = g_D^2\beta^2 \sim (68.5\beta)^2$.
This energy loss mechanism is implemented as part of a GEANT package \cite{geant} for the simulation of monopole trajectories in a detector \cite{bauer}.

\begin{figure*}[tb]
%\vspace*{2mm}
\begin{center}
\includegraphics[width=16.0cm]{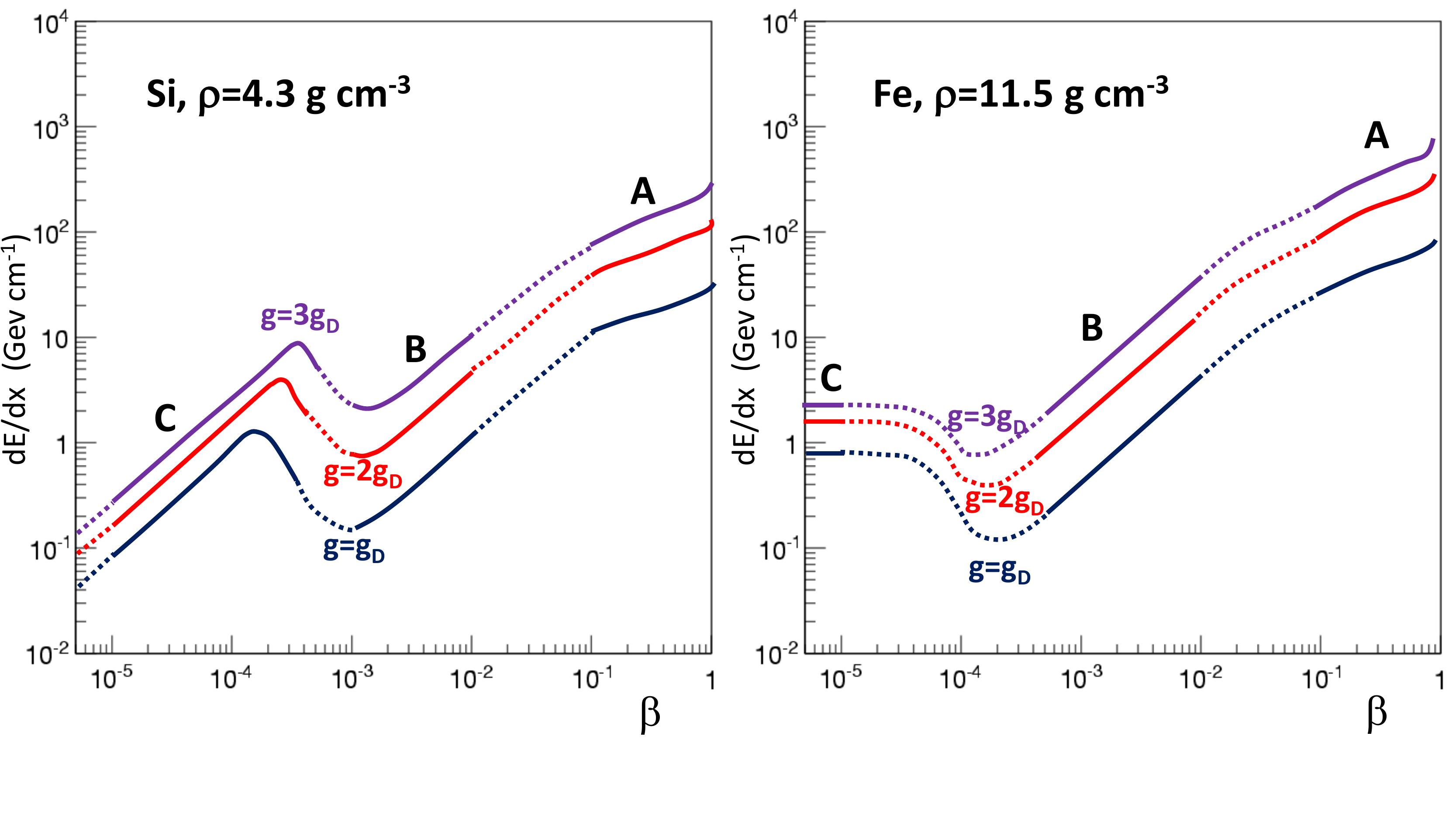}
\end{center}
\vspace*{-12mm}
\caption{\small\label{fig:eloss} a) Energy losses of $\cm$s in the Earth's mantle and b) in the Earth's nucleus as function of the speed $\beta $ for different values of the magnetic charge $g$. Notice the solid curves in the regions A, B, C where the calculations are more reliable, and the dashed lines when they are interpolated. The three curves apply to $\cm$s with $n = 1,2,3$ unit Dirac charge and assuming that the basic electric charge is that of the electron.
The curves labeled $g=3g_D$ apply also to $\cm$ with $g=g_D$ assuming that the basic electric charge is that of the quark $d$. }
\end{figure*}

Fig. \ref{fig:eloss} shows the $\cm$ stopping power in silicon (a) at the density of the Earth's mantle, and in iron (b) at the density of the Earth's core. The energy loss computed according to Eq. \ref{eq:3.elos} is labeled with A.

%\paragraph{Very large velocities} 
%%%%%%%%%%%%%%%%%%%%%%%%%%%%%%%%%
When $\beta \gamma> 10^4$ stochastic processes such as bremsstrahlung, pair production and hadron production dominate the energy loss. 
Photonuclear processes yield the largest contribution in materials with moderate values of atomic mass (such as rock or water), with a smaller contribution from pair production. 
In contrast pair production is the dominant process in materials with high atomic number $Z$.
The contribution from bremsstrahlung is small and is usually neglected.
Theoretical uncertainties arising from the large coupling constant $\alpha_m$ should be considered in the computation of transition probabilities.
Thus, the energy loss per unit path length in this energy regime is dominated by single (or small numbers) of collisions with the possibility of a large energy transfer, leading to large fluctuations on the stopping power and range of the particle.

%%%%%%%%% 
%%%%%%%%% 

%%%%%%%%%%%%%%%% 
\subsection{Stopping power of monopoles with $10^{-3}\lesssim \beta \lesssim 10^{-2}$ }
\label{sec:3-eloss-2}
%%%%%%%%%%%%%%%%
This velocity range is of particular interest because of the considerations given in \S \ref{sec:GUTMM} for massive $\cm$s bound to astrophysical systems.
In the calculation of the energy loss the materials is approximated by a free (degenerate) gas of electrons. 
This is appropriate in the description of $\cm$'s interactions with the conduction electrons of metallic absorbers.
For nonmetallic absorbers it represents a reasonable approximation for heavy atoms $(Z \ge 10)$.  
Using this approximation, Ahlen and Kinoshita \cite{ak82} have computed the stopping power in the velocity range $10^{-3}\lesssim \beta \lesssim 10^{-2}$ as:
\begin{equation}\label{eq:3.ak}
{dE\over dx} = {2\pi N_e g^2 e^2 \beta \over m_e c v_F} \biggl[ \ln {2m_ev_F a_o \over \hbar} - {1\over 2} \biggr] \ ,
\end{equation}
where $v_F = (\hbar/m_e) (3\pi^2N_e)^{1/3} \simeq 3.9 \times 10^8$ cm/s is the Fermi velocity (the value applies to silicon at the Earth's mantle density, $\rho_{Si} = 4.3$ g/cm$^3$); $a_o$ is the Bohr radius; $N_e$ is the density of electrons.

The expected $\cm$ energy loss is larger than that computed with Eq. \ref{eq:3.ak} if the contribution from the coupling of the magnetic monopole with the electron magnetic moment is considered. 
According to \cite{ak82} this gives a multiplicative factor of 1.37 to Eq. \ref{eq:3.ak}. 
For conductors one should add another term that depends on the conduction electrons. 
In Fig. \ref{fig:eloss} regions labeled with B indicate the expected stopping power of $\cm$ in the Earth's mantle (a) and core (b) for $10^{-3}\lesssim \beta \lesssim 10^{-2}$.

Note that the stopping power mechanism parameterized with Eq. \ref{eq:3.ak} refers to an average energy transferred to atomic electrons so small that it is impossible to disregard their atomic bindings.
In addition, transitions yielding photons in the visible range are largely suppressed. For this reason, as pointed out by the authors \cite{ak82}, Eq. \ref{eq:3.ak} cannot be directly used for the evaluation of the response of excitation- and ionization-sensitive particle detectors to slow, supermassive $\cm$s.

%%%%%%%%%%%%%%%% 
\subsection{Stopping power of monopoles with $ \beta \lesssim 10^{-3}$ }
\label{sec:3-eloss-3}
%%%%%%%%%%%%%%%% 

In this velocity range $\cm$s cannot excite atoms but they can lose energy in elastic collisions with atoms or nuclei through the $\cm$'s coupling with the atomic/nuclear magnetic moment. 
In \cite{derk98} different computations were done for diamagnetic and paramagnetic materials. 
Fig. \ref{fig:eloss} (solid lines C) shows the stopping power in the Earth's mantle and nucleus for slowly moving $\cm$s. 
Dashed lines between solid lines B and C are interpolations; dashed lines at $\beta < 10^{-5}$ are extrapolations.
At low velocities, the energy loss is mainly released to the medium in the form of thermal and acoustic energy. 
Particle detectors are sensitive to $\cm$s only through particular interaction mechanisms, as the Drell effect (see \S \ref{sec:4-gas}), and via atomic recoils.

%%%%%%%%% 
\begin{figure*}[tb]
%\vspace*{2mm}
\begin{center}
\includegraphics[width=11.5cm]{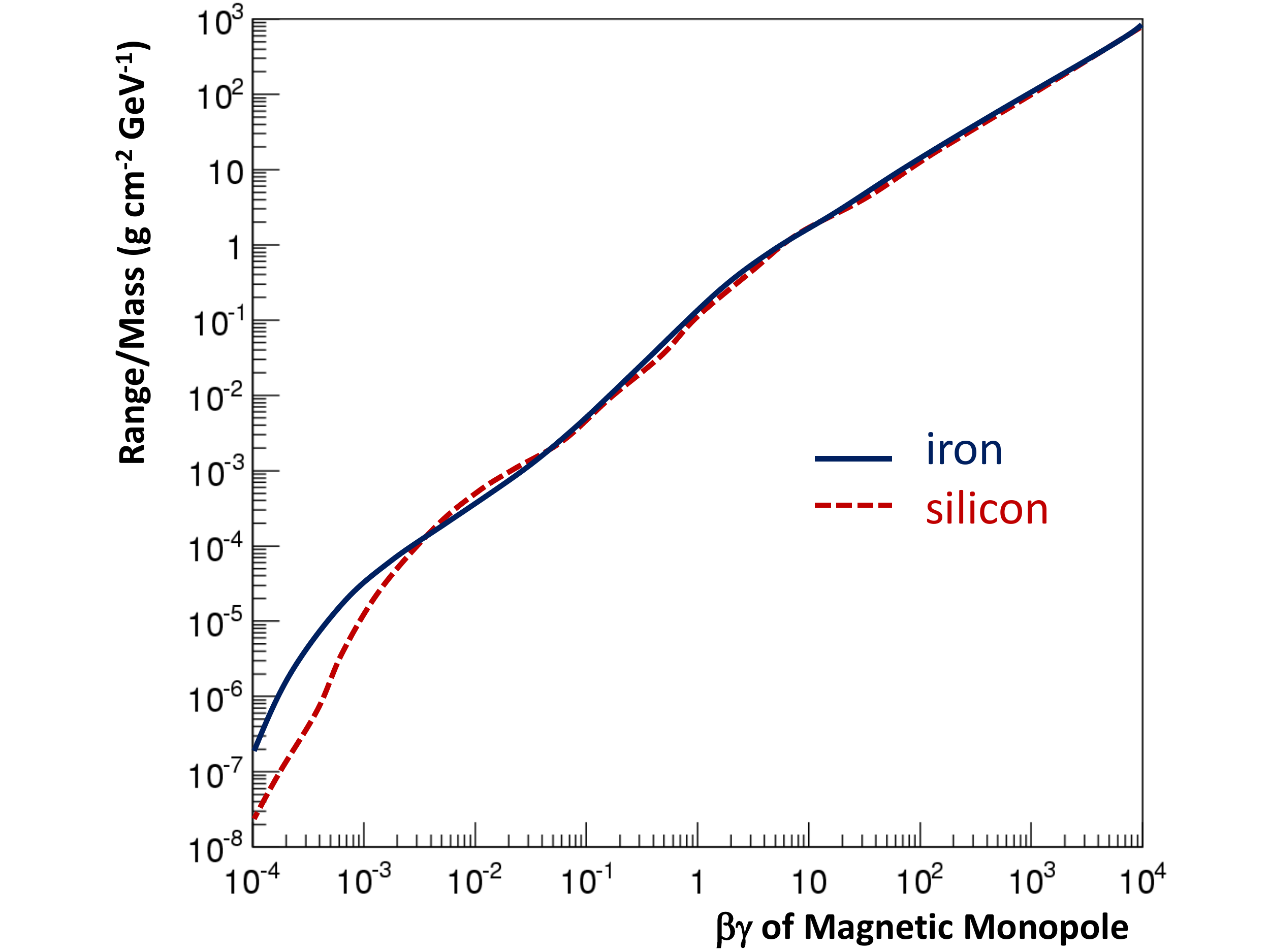}
\end{center}
\vspace*{-2mm}
\caption{\small\label{fig:rm} 
The ratio range/mass for $\cm$s with $g_D$ magnetic charge in iron and silicon as a function of monopole $\beta\gamma=p/M$. The range is computed from the stopping powers given in \cite{derk98} and it is defined as the thickness of material to slow down the monopole to $\beta = 10^{-6}$.}
\end{figure*}
%%%%%%%%% 

%%%%%%%%%%%%%%%%%%%%%%%%%%%%%%%%%%%%%%%%%%%%%%%%
\subsection{Range of magnetic monopoles\label{sec:3-range} }
% arxiv 1410.1374
%%%%%%%%%%%%%%%%%%%%%%%%%%%%%%%%%%%%%%%%%%%%%%%%
Magnetic monopoles can cross the atmosphere and reach underground detectors depending on their stopping power, which in turn depends on the monopole mass $M$ and velocity.
The range $R$ can be computed by integrating the stopping power:
\begin{equation}\label{eq:3.range}
R=\int_{E_{min}}^{E_0} {dE\over dE/dx} \ ,
\end{equation}
where $E_0$ is the pole initial kinetic energy, and $E_{min}$ the kinetic energy corresponding to $\beta \simeq 10^{-6}$, when it can be considered as stopped. 
Fig. \ref{fig:rm} shows the ratio $R/M$ (in g cm$^{-2}$ GeV$^{-1}$), as a function of the initial $\beta\gamma=p/M$ of the $\cm$, with $p$ being its momentum \cite{burdin}. The stopping power used in the computation is that obtained in Si and Fe in \cite{derk98}.  

Since the range depends on the kinetic energy, the $R/M$ ratio is independent of the mass.
From Fig. \ref{fig:rm}, we can estimate that the minimum mass for, e.g.,  a $\beta=10^{-3}$ magnetic monopole to cross the Earth atmosphere (1000 g cm$^{-2}$) is $M\simeq 3\times 10^7$ GeV. It increases to about $M\simeq 10^{10}$ GeV for a monopole penetrating $3\times 10^5$ g cm$^{-2}$, corresponding to the minimum rock overburden of the Gran Sasso laboratory.

%%%%%%%%%%%%%%%%%%%%%%%%%%%%%%%%%%%%%%%%%%%%%%%%%%%%%%%%%%%
%%%%%%%%%%%%%%%%%%%%%%%%%%%%%%%%%%%%%%%%%%%%%%%%%%%%%%%%%%%
%%%%%%%%%%%%%%%%%%%%%%%%%%%%%%%%%%%%%%%%%%%%%%%%%%%%%%%%%%%
%%%%%%%%%%%%%%%%%%%%%%%%%%%%%%%%%%%%%%%%%%%%%%%%%%%%%%%%%%%
%%%%%%%%%%%%%%%%%%%%%%%%%%%%%%%%%%%%%%%%%%%%%%%%%%%%%%%%%%%
\section{Experimental Methods for Magnetic Monopole Searches}
\label{sec:4-tech}
%%%%%%%%%%%%%%%%%%%%%%%%%%%%%%%%%%%%%%%%%%%%%%%%%%%%%%%%%%%
%%%%%%%%%%%%%%%%%%%%%%%%%%%%%%%%%%%%%%%%%%%%%%%%%%%%%%%%%%%
%%%%%%%%%%%%%%%%%%%%%%%%%%%%%%%%%%%%%%%%%%%%%%%%%%%%%%%%%%%
%%%%%%%%%%%%%%%%%%%%%%%%%%%%%%%%%%%%%%%%%%%%%%%%%%%%%%%%%%%
%%%%%%%%%%%%%%%%%%%%%%%%%%%%%%%%%%%%%%%%%%%%%%%%%%%%%%%%%%%

Search strategies for magnetic monopoles are determined by their expected interactions as they pass through a detector. 
Searches in cosmic rays, in matter, and at accelerator experiments are based on different techniques, as the induction method, exploiting the electromagnetic interaction of the $\cm$ with the quantum state of a superconducting ring, or the $\cm$'s energy loss in different detectors.

%%%%%%%%%%%%%%%%%%%%%%%%%%%%%%%%%%%%%%%%%%%%%%%%%%%%%%%%%%%
\subsection{The induction technique in superconductive coils}
\label{sec:4-super}
%%%%%%%%%%%%%%%%%%%%%%%%%%%%%%%%%%%%%%%%%%%%%%%%%%%%%%%%%%%

The technique of looking for monopoles using small superconducting coils started in mid 1970s \cite{75E2}.
The detection method consists in a superconducting coil coupled to a SQUID (Superconducting Quantum Interferometer Device). 
A moving $\cm$ induces in a superconducting ring an electromotive force and a current change $(\Delta i)$. For a coil with $N$ turns and inductance $L$ the current change is $\Delta i = 4\pi N n g_D/L = 2\Delta i_o$, where $i_o$ is the current change corresponding to a change of one unit of the flux quantum of superconductivity. 
This method is based only on the long-range electromagnetic interaction between the magnetic charge and the macroscopic quantum state of a superconducting ring.
For any sample of matter containing only magnetic dipoles, induced currents cancel out to zero after its passage through the coil.
A persistent current directly proportional to $g_D$ would represent an unmistakable signature of the $\cm$'s passage. 
Induction coils are the only known device sensitive only to the magnetic charge, so able to detect poles of any velocity and mass.
The system components, in particular the magnetometer, have to be extremely well shielded from any variation of the ambient magnetic field. This places severe restrictions on the cross-sectional area of induction detectors. 

SQUID have been used in the past to search for $\cm$s in the cosmic radiation. 
In 1982 a group in Stanford operated a four-turn coil of 5 cm diameter for 151 days \cite{cabrera}, recording a single current jump corresponding to that expected from a magnetic charge equal to $g_D$.
No other jumps were observed in subsequent runs. 
That candidate event generated a great deal of interest on $\cm$s, and the induction detection technique evolved quickly toward second-generation experiments. 
No additional candidates were found. The technique has not been used anymore to search for $\cm$s in the cosmic radiation, while it is still widely used in searches for $\cm$s trapped in matter, \S \ref{sec:5-bouma}.

%%%%%%%%%%%%%%%%%%%%%%%%%%%%%%%%%%%%%%%%%%%%%%%%%%%%%%%%%%%
\subsection{Light yield in scintillators}
\label{sec:4-scint}
%%% Derk99
%%%%%%%%%%%%%%%%%%%%%%%%%%%%%%%%%%%%%%%%%%%%%%%%%%%%%%%%%%%

Different computations exist for the light yield of $\cm$s and dyons in a liquid or plastic scintillator. For a detailed review, see \cite{derk99}.
The light yield $(dL/dx)$ in a scintillator is related to the total electronic energy loss $(dE/dx)$ by 
\begin{equation}\label{eq:4.dldx}
{dL\over dx} = {\cal A} \times \biggl[ { 1-F\over 1+{\cal AB}(1-F) {dE\over dx} } + F   \biggr] \times  {dE\over dx} \ .
\end{equation}
For relatively small energy losses, $dL/dx\simeq {\cal A}\times dE/dx$, with ${\cal A}\simeq 0.03\div 0.07$, depending on the material and on $\cm$'s velocity. 
The light yield from the energy deposited near the particle track (the first term in the parenthesis of Eq. \ref{eq:4.dldx}) saturates for high energy losses. 
The term ${\cal B}=0.3\div 0.7$ cm/MeV is called quenching parameter. 
The term $F$ (which is small in most cases) represents the fraction of energy loss which results from excitations outside the track core; these excitations are assumed to be not quenched.

\begin{figure*}[tb]
\begin{center}
\includegraphics[width=16.0cm]{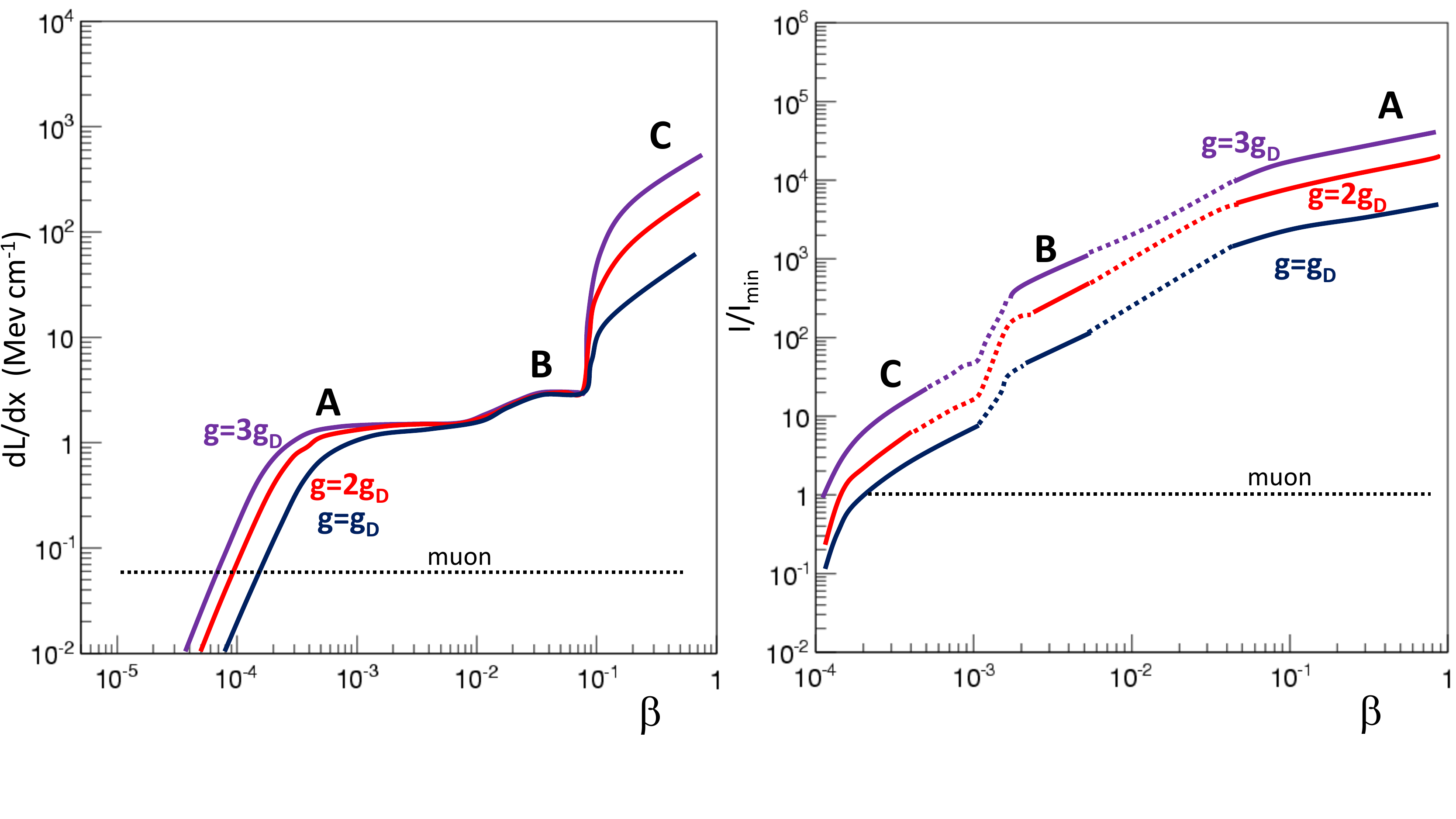}
\end{center}
\vspace*{-12mm}
\caption{\small\label{fig:lst} a) Light yield of $\cm$s in plastic or liquid scintillators as a function of the speed $\beta$ for different magnetic charges $g$. 
The light yield from a minimum ionizing muon is shown for comparison. 
b) Energy losses produced in the limited streamer tubes filled with 73\% He and 27\% n-pentane, $\rho = 0.856\times 10^{-3}$ g/cm$^3$, by $\cm$s as a function of $\beta$, relative to the ionization produced by a minimum ionizing particle, $I_{min} = 2.2$ MeV g/cm$^2$. See text for the explanation of labels A, B and C. Notice that the solid curves represent the regions where the calculations are more reliable; the dashed lines are interpolated values.  }
\end{figure*}

The results of detailed calculations in a wide range of velocities is shown in Fig. \ref{fig:lst}a).
We notice that at very low velocities the light yield increases with $\beta$ then it saturates at $\sim$ 1.2 MeV/cm (region A).
The increase in the light yield observed in region B is due to changes in the quenching parameter.
For $\beta\ge 0.09$ some ionized electrons have sufficient energy to escape from the region near the track core ($\delta$-rays), and the light yield increases again with $\beta$ (region C).
At intermediate $\beta$ values ($0.003 < \beta < 0.1)$ - curve B - the light yield is almost independent of the magnetic charge $g$; at low and high $\beta$ the light yield increases as $g^2$.
The light yield at low velocities has to be considered a lower limit, since the calculations do not take into account possible contributions that can arise from the mixing and crossing of molecular electronic energy levels at the passage of the magnetic charge. This could result in molecular excitations and emission of additional light.

%%%%%%%%% 
%%%%%%%%% 

%%%%%%%%%%%%%%%%%%%%%%%%%%%%%%%%%%%%%%%%%%%%%%%%%%%%%%%%%%%
\subsection{Ionization in gaseous detectors and the Drell effect}
\label{sec:4-gas}
%% APP_Scin
%%%%%%%%%%%%%%%%%%%%%%%%%%%%%%%%%%%%%%%%%%%%%%%%%%%%%%%%%%%

A gaseous detector (e.g., the drift tubes of the Soudan 2 experiment or the streamer tube sub-system of the MACRO experiment, see \S \ref{sec:5-cosmic}) is characterized by a pressure corresponding about to that of the atmosphere. Thus, the resulting density is very low (in comparison with the density of other detectors), order of $\sim 10^{-3}$ g/cm$^3$.

For velocities $\beta > 0.05$ the monopole energy loss in gaseous detectors is well described by Eq. \ref{eq:3.elos}. 
The ionization potential of the medium is $\sim 50$ eV, the resulting energy losses are given in Fig. \ref{fig:lst}b), by curves labeled A.

At intermediate velocities, $10^{-3} < \beta < 10^{-2}$, the usual approach is to consider the medium as a degenerate electron gas, \S \ref{sec:3-eloss-2}. 
At the low densities of a gas, the electrons are not free, but bound in well-separated molecules. Thus, the assumptions yielding the curves B in Fig. \ref{fig:lst}b) are affected by larger uncertainties. 
The dashed lines indicate regions where the calculations are less reliable than those for the solid lines.

%%%%%%%%%%%%%%%%%%%%%%%%%%%%%%%%%%%%%%%%%%%%%%%%
%\subsection{The Drell effect\label{sec:3-drell} }
%%%%%%%%%%%%%%%%%%%%%%%%%%%%%%%%%%%%%%%%%%%%%%%%

%\paragraph{The Drell effect\label{sec:3-drell}}
%%%%%%%%%%%%%%%%%%%%%%%%%%%%%%%%%%%%%%%%%%%%%%%%%%%%%
At low velocities, $10^{-4} < \beta < 10^{-3}$, and for particular gases as hydrogen and helium, the Drell effect \cite{drell} occurs.
Due to the coupling between the pole magnetic field and the electron magnetic moment, the atomic energy levels are changed by the passage of a slowly moving magnetic charge, and an electron can make a transition to an excited level. 
The energy levels of the atom split in the characteristic Zeeman pattern. The $\cm$'s energy loss can be observed in the form of subsequent electromagnetic radiation when the excited electrons return to their ground states. 
The Drell mechanism is effective as long as the monopole-atom collision
energy exceeds the spacing of atomic levels. 
The effect may be used for practical detection either by observing photons emitted in the de-excitation of atoms or by observing the ionization caused by the energy transfer from the excited atoms to complex molecules with a low ionization potential (the Penning effect).
Helium plus n-pentane (C$_5$H$_{12}$) was used for this purpose in the MACRO streamer tube sub-system, \S \ref{sec:5-macro}.
The results of the calculations performed in \cite{derk99} for different values of the magnetic charge $g$ are shown in Fig. \ref{fig:lst}b), by curves labeled with C.

%%%%%%%%%%%%%%%%%%%%%%%%%%%%%%%%%%%%%%%%%%%%%%%%%%%%%%%%%%%
\subsection{Energy loss mechanisms in nuclear track detectors}
\label{sec:4-ntd}
%%%%%%%%%%%%%%%%%%%%%%%%%%%%%%%%%%%%%%%%%%%%%%%%%%%%%%%%%%%

\begin{figure*}[tb]
%\vspace*{2mm}
\begin{center}
\includegraphics[width=13.0cm]{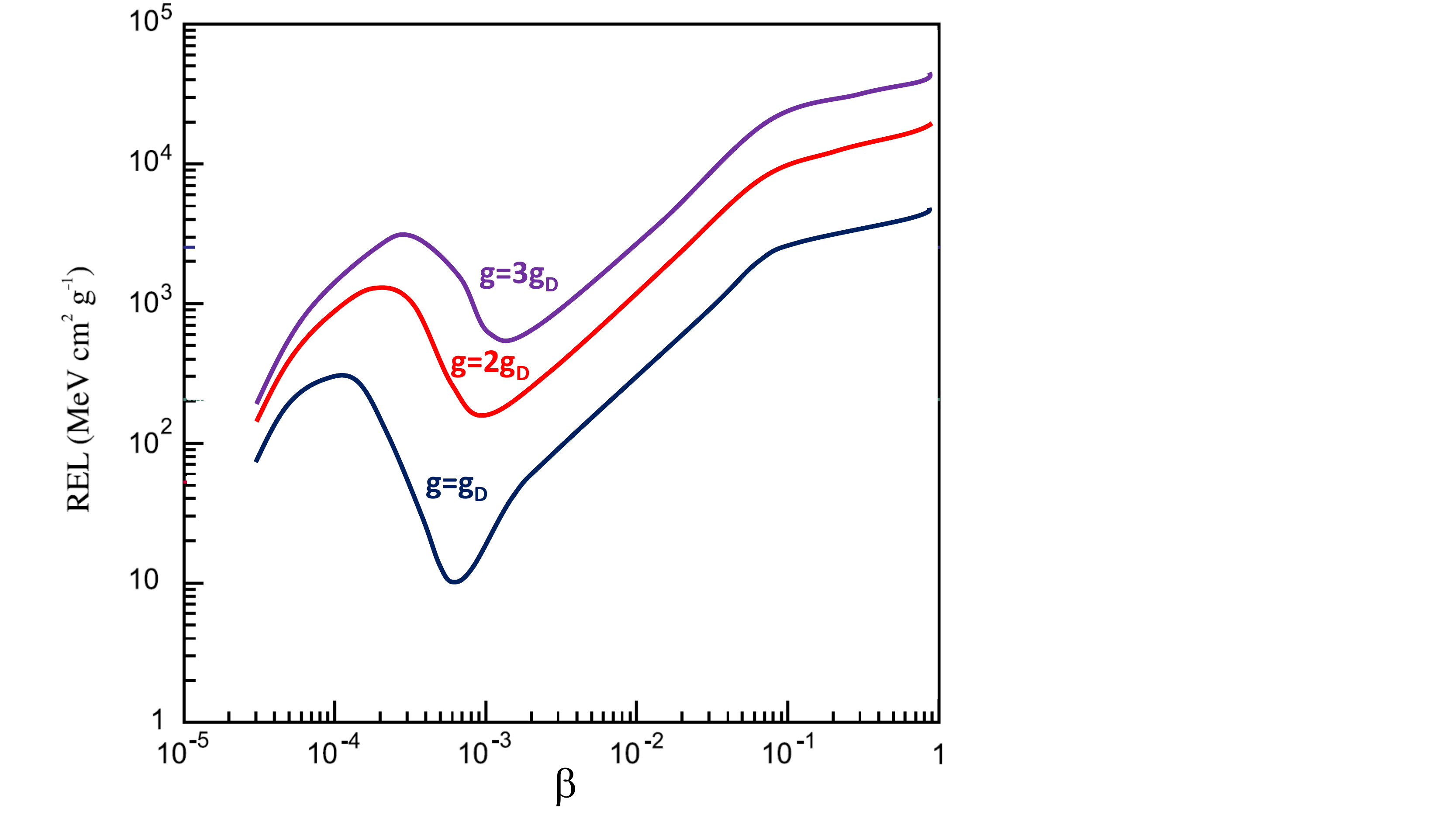}
\end{center}
\vspace*{-6mm}
\caption{\small\label{fig:rel}
Restricted Energy Losses (REL) as a function of $\beta$ for $\cm$s with magnetic charges $g = g_D, 2g_D$, and $3g_D$.} 
\end{figure*}

The passage of heavily ionizing particles may be permanently recorded as a \textit{latent track} in some insulating materials (called nuclear track detectors, NTDs) as polymers, kapton and nitrocellulose, to glasses and to minerals like mica and obsidian. 
NTDs are sensitive to the fraction of the energy loss released in a narrow region close to the particle's trajectory.
In polymers like CR39{\textregistered} or Makrofol{\textregistered}/Lexan\texttrademark the restricted energy loss (REL) is the relevant quantity. 
%The REL includes contribution from $\delta$-rays with short range.
At high $\cm$ velocities ($\beta > 0.05$) the REL excludes the energy transfers resulting in energetic $\delta$-rays and thus in energy deposited far away from the track \cite{ah88}. 
At lower $\cm$ velocities $(\beta < 10^{-2}$), elastic recoils from diamagnetic interactions between the $\cm$ and the atoms have to be considered. 
% due to the diamagnetic interaction between the $\cm$ and the atoms (for %instance carbon and oxygen) must be also considered. 
A detailed computation of these energy loss mechanisms are in \cite{derk99} and the REL as a function of $\beta$ is shown in Fig. \ref{fig:rel}. The atomic elastic recoil contribution gives rise to a bump in REL below $\beta \sim 10^{-3}$. 

The latent track produced by the REL may be made visible under an optical microscope by proper chemical etching \cite{gg84}. 
NTDs are detection threshold devices, with no time resolution. The threshold depends on the material and on the chemical etching.
By chemical etching, two regular etch-pit cones are formed on each side of the detector sheet along the track of a crossing particle. 
The REL may be determined from the geometry of the etched cones. 
For CR39 this technique is particularly successful, yielding measurements of the electric charge of heavy relativistic nuclei ($Z>5$) to a precision of 0.05$e$ if several layers of plastic sheets are used \cite{cr39-1,cr39-2}. 

The CR39 NTD was calibrated with relativistic and low velocity ($\beta \ge 4\times 10^{-3}$) ions over the entire range of REL relevant for GUT $\cm$'s \cite{cr39NC}.
The REL mechanism describes the polymer response in the whole interval of velocity; in particular, below the excitation-ionization regime, data are consistent with about 100\% contribution of the atomic elastic recoil.
This measurement represents an experimental demonstration that the CR39 can effectively record the passage of a $\cm$ through the formation of the latent track due to induced atomic elastic recoils.

%%%%%%%%% 
%%%%%%%%% 

%%%%%%%%%%%%%%%%%%%%%%%%%%%%%%%%%%%%%%%%%%%%%%%%%%%%%%
%%%%%%%%%%%%%%%%%%%%%%%%%%%%%%%%%%%%%%%%%%%%%%%%%%%%%%
%%%%%%%%%%%%%%%%%%%%%%%%%%%%%%%%%%%%%%%%%%%%%%%%%%%%%%
%%%%%%%%%%%%%%%%%%%%%%%%%%%%%%%%%%%%%%%%%%%%%%%%%%%%%%
\section{Searches for Magnetic Monopoles}
%\vspace{-0.2cm}{\it\small( 9 pages)}
\label{sec:searches}
%%%%%%%%%%%%%%%%%%%%%%%%%%%%%%%%%%%%%%%%%%%%%%%%%%%%%%
%%%%%%%%%%%%%%%%%%%%%%%%%%%%%%%%%%%%%%%%%%%%%%%%%%%%%%
%%%%%%%%%%%%%%%%%%%%%%%%%%%%%%%%%%%%%%%%%%%%%%%%%%%%%%'
%%%%%%%%%%%%%%%%%%%%%%%%%%%%%%%%%%%%%%%%%%%%%%%%%%%%%%

Searches for $\cm$s have been performed at accelerators, in cosmic rays, and trapped in matter. 
No confirmed observations of exotic particles possessing magnetic charge exist. 
The limits on the flux of cosmic $\cm$ or on the production cross sections at accelerators are usually made under the assumption of bare magnetic charge.
Most of the searches are also sensitive to dyons; on the contrary, dedicated analyses are required to search for magnetic monopoles inducing the catalysis of proton decay.

An extensive bibliography on magnetic monopoles till mid of 2011 is in \cite{GGbio1,GGbio2}. It was intended to contain nearly all the experimental papers on the subject and only the theoretical papers, which have some specific experimental implications.

%%%%%%%%%%%%%%%%%%%%%%%%%%%%%%%%%%%%%%%%%%%%%%%%%%%%%%
\subsection{Searches at colliders \label{sec:5-class}}
%%%%%%%%%%%%%%%%%%%%%%%%%%%%%%%%%%%%%%%%%%%%%%%%%%%%%%
The possible production of low mass Dirac $\cm$s has been investigated at $e^{+}e^{-} $, $e^{+}p$, $p\overline{p}$ and $pp$ colliders, mostly using scintillation counters, wire chambers and nuclear track detectors \cite{fair07}. 
Searches for monopoles trapped in the material surrounding the collision regions were also made.

\begin{figure*}[tb]
%\vspace*{2mm}
\begin{center}
\includegraphics[width=12.0cm]{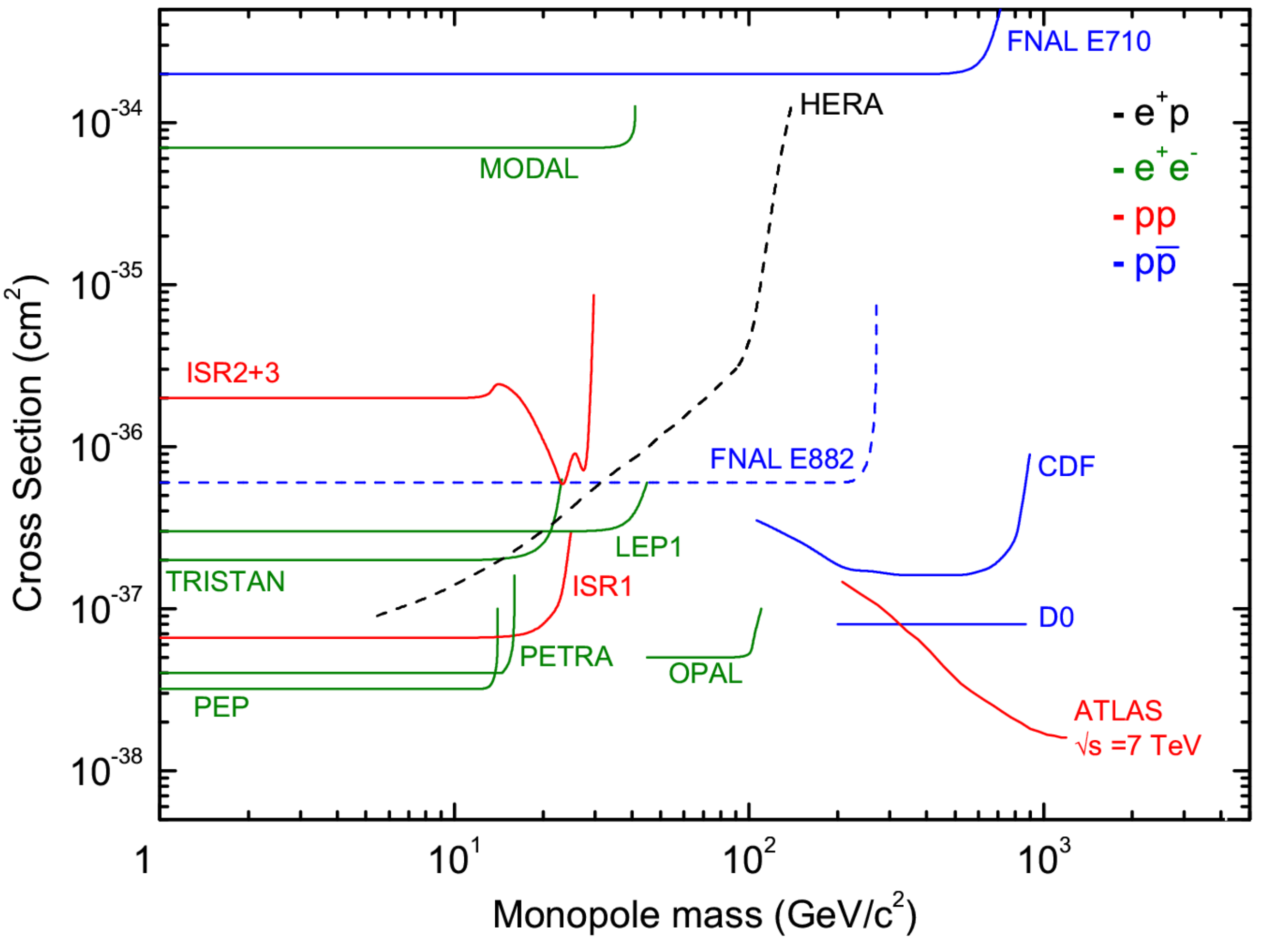}
\end{center}
\caption{\small\label{fig:acce} Cross section upper limits on monopole production from searches at colliders. All limits are at 95\% CL, besides E882 given at 90\% CL.  
Dashed lines indicate limits from indirect searches of $\cm$ trapped in the beam pipe/detector materials.}
\end{figure*}

The searches have set upper limits on $\cm$ production cross sections for masses below the TeV scale, as reported in Fig. \ref{fig:acce}.
These limits, based on assumptions of the production processes of monopole-antimonopole pairs, are model-dependent.

%\paragraph{Searches at the $pp$ collider}
%%%%%%%%%%%%%%%%%%%%%%%%%%%%%%%%%%%
The highest energies are available in $pp$ collisions at the CERN LHC at a center-of-mass energy $\sqrt{s}=7$ TeV. Searches for highly ionizing particles leaving characteristic energy deposition profiles were sought by the ATLAS experiment \cite{atlas}, with no event found. The corresponding upper limit on the cross section at 95\% confidence level (CL) varies from 145 fb to 16 fb for $\cm$s with mass between 200 GeV and 1200 GeV.
Early experiments at the CERN ISR \cite{isr1,isr23} reported limits in the mass range $\lesssim 30$ GeV.

%\paragraph {Searches at $e^+e^-$ colliders}
%%%%%%%%%%%%%%%%%%%%%%%%%%%%%%%%%%%
Searches at $e^{+}e^{-}$ colliders were based on the detection of $\cm$ pairs produced through the $e^{+}e^{-} \rightarrow \gamma^{*} \rightarrow \cm \overline{\cm}$ reaction.
The search with the OPAL detector \cite{OPAL} at the CERN LEP used data collected at $\sqrt{s}=206.3$ GeV (with a total integrated luminosity of $62.7$ pb$^{-1}$) and it was based on the measurements of the momentum and energy loss in the tracking chambers of the detector. 
Back-to-back tracks with high-energy release were searched for in opposite sectors of the Jet Chamber. The 95\% CL upper limit for the production of monopoles with masses $45 <{M}<104$ GeV was set at $0.05$ pb.
The MODAL experiment \cite{modal} collected data at $\sqrt{s} = 91.1$ GeV and integrated luminosity of $\sim 60$ nb$^{-1}$.  
The detector consisted of a polyhedral array of CR39 NTD foils covering a $0.86 \times 4\pi$ sr angle surrounding the I5 interaction point at LEP.
%After chemical etching NTD sheets were analyzed searching for %penetrating tracks consistent with the passage of a heavily ionizing %particle. No candidate event was found; the 95\% CL upper limit on the %$\cm$ production cross section was set at $70$ pb for monopoles with %masses $<45$ GeV.
Another experiment (LEP1 in Fig. \ref{fig:acce}) searching for highly ionizing particles produced at the OPAL intersection point used Lexan NTD\cite{lep1}.
NTDs around the beam interaction point were also used in earlier searches at $e^+e^-$ colliders such as the TRISTAN ring at KEK \cite{kek}, PETRA at DESY \cite{petra}, and PEP at SLAC \cite{pep}.

%\paragraph {Searches at $p\overline p$ collider}
%%%%%%%%%%%%%%%%%%%%%%%%%%%%%%%%%%%%%%%%%%
At the FNAL Tevatron $\overline{p}p$ collider, the CDF collaboration \cite{CDF} performed a search for $\cm$s produced in 35.7 pb$^{-1}$ integrated luminosity at $\sqrt{s}= 1.96$ TeV.
$\cm$s would have been detected by the drift chambers constituting the Central Outer Tracker, and by the Time of Flight scintillator detectors placed in the 1.4 T magnetic field parallel to the beam direction. The ${\cm \overline{\cm}}$ pair production was excluded at 95\% CL for cross sections larger than $0.2$ pb and monopole masses $200<{M}< 700$ GeV.
The D0 collaboration \cite{D0} performed a search for heavy point-like Dirac monopoles by searching for pairs of photons with high transverse energies. Pairs of nearly real photons can be produced by $\cm$s via a particular box Feynman diagram.
Finally, NTDs were used by the FNAL E710 experiment \cite{bertani}.

%%%%%%%%% 
%%%%%%%%% 

If produced in high-energy collisions, $\cm$s could stop in material surrounding the interaction points of a collider. 
Discarded parts of the beam pipe or detector material near the interaction regions can be divided into small pieces and passed through a superconducting coil coupled to a SQUID; the signature of the presence of a $\cm$ would be the induction of a persistent current in the superconducting loop \cite{Milton}. 

This method was applied by the H1 collaboration \cite{hera} using the aluminum beam pipe of the HERA lepton-proton collider, exposed to a luminosity of $62\pm 1$ pb$^{-1}$ at $\sqrt{s} = 300$ GeV, and by the E882 collaboration \cite{E882}, which looked for monopoles trapped in the material surrounding the D0 and CDF  collision regions.
An initial, preliminary search has already been done using a limited set of available samples.
The main purpose of this research \cite{squidLHC} was to quantify the expected response of the magnetometer to monopoles, study how fake signals can arise, and run through the protocols needed for a sample release from CERN. A much better-performing search will be carried out when a much larger integrated luminosity at the LHC is collected for material exposed to pp and heavy-ion collisions.

%%%%%%%%%%%%%%%%%%%%%%%%%%%%%%%%%%%%%%%%%%%%%%%%%%%%%%
\subsection{Direct searches in the Cosmic Radiation}\label{sec:5-cosmic}
%%%%%%%%%%%%%%%%%%%%%%%%%%%%%%%%%%%%%%%%%%%%%%%%%%%%%%
%%%%%%%%%%%%%%%%%%%%%%%%%%%%%%%%%%%%%%%%%%%%%%%%%%%% 
\begin{figure*}[tb]
%\vspace*{2mm}
\begin{center}
\includegraphics[width=12.0cm]{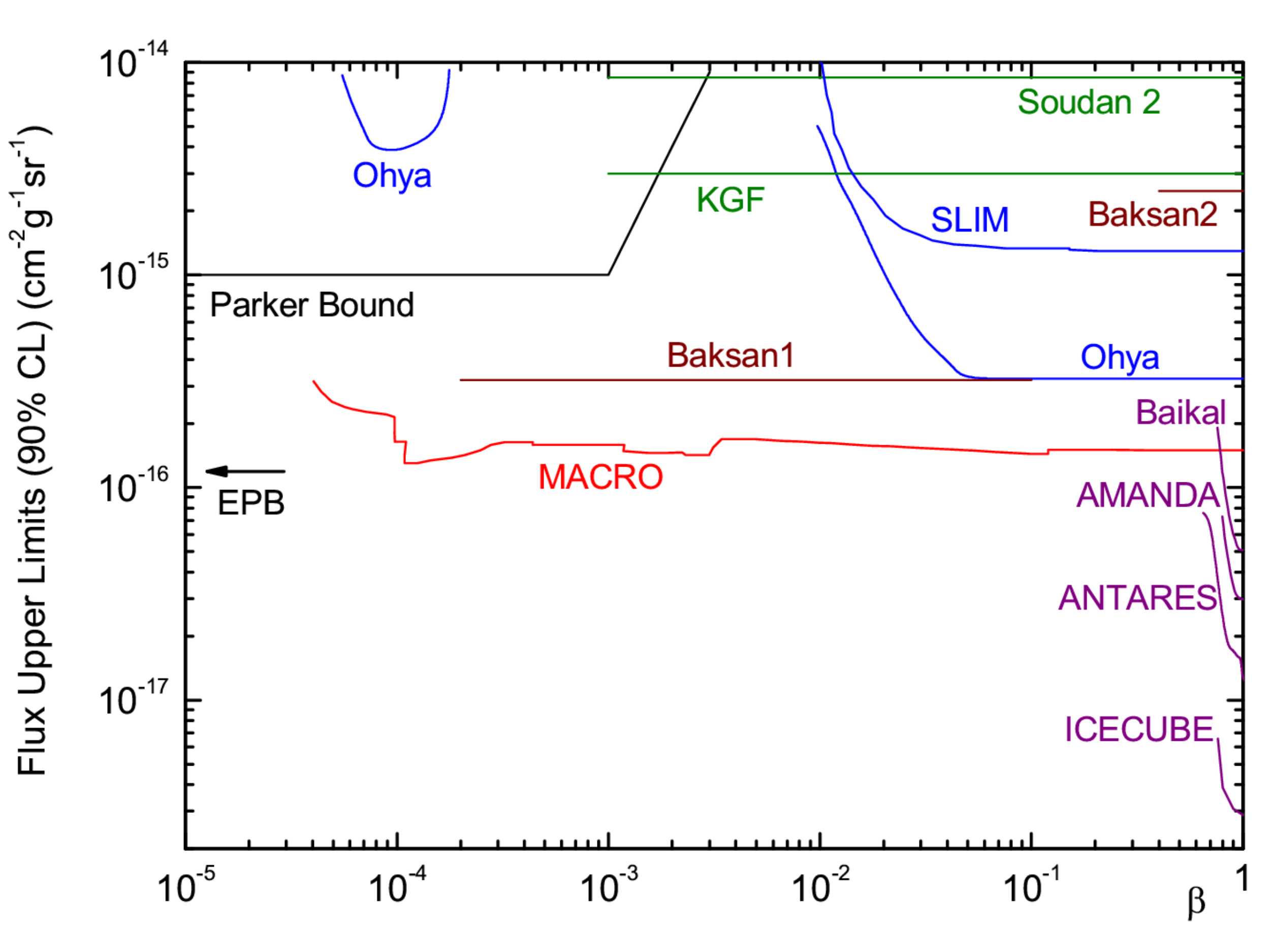}
\end{center}
\caption{\small\label{fig:under} The 90\% CL upper limits vs $\beta$ for a flux of cosmic GUT monopoles with magnetic charge $g=g_{D}$.
The Parker bound refers to Eq. \ref{eq:2.din6}. while the extended Parker bound (EPB) refers to Eq. \ref{eq:2.din7}.}
\end{figure*}
%%%%%%%%%%%%%%%%%%%%%%%%%%%%%%%%%%%%%%%%%%%%%%%%%%%% 

%\paragraph{Excitation/ionization methods}
%%%%%%%%%%%%%%%%%%%%%%%%%%%%%%%%%%%%%%%%%%%%%%%%%%%%%%%%
Direct searches for $\cm$s in cosmic rays refer to experiments in which the passage of particles is recorded by detectors under controlled conditions.
Several searches for GUT $\cm$s were performed above ground and underground using different types of detectors \cite{gps}. 
$\cm$s with masses ${M}>10^{5}-10^{7}$ GeV in the range of $\beta \sim 10^{-3}-10^{-1}$ could reach the Earth surface from above and be detected, as discussed in \S \ref{sec:3-range}.  
Lower mass $\cm$s may be searched for with detectors located at high mountain altitudes, balloons and satellites.

The background such searches have to fight with is mainly due to muons of the cosmic radiation \cite{cecco} and natural radioactivity. 
For this reason, different large detectors have been installed in underground laboratories. 
The minimum mass of $\cm$s to reach underground detectors from above (at a given velocity) depends on the overburden of the experiment \cite{slim}.

Magnetic monopoles can be detected in the velocity range $10^{-3}\le \beta<1$ exploiting their large ionization energy loss either in gaseous detectors, in scintillation counters, and NTDs.
In the velocity range $10^{-4}<\beta<10^{-3}$ monopoles do not induce  ionization in matter (besides via the Drell-Penning effect), although they may still excite atoms which would emit scintillating light.
In all searches, the cross section for the catalysis process is typically set to very low values \cite{02A3}, equivalent to a null effect in the detector. 
Searches exploiting the expected catalyzed decays are discussed in \S \ref{sec:5-cata}.

The most stringent flux upper limits on supermassive $\cm$s in the widest $\beta$ range obtained using different experimental techniques were set by the MACRO experiment at the underground Gran Sasso laboratory (Italy) \cite{02A3}.
This experiment represents a benchmark for magnetic monopole searches, and for this reason, it deserves a dedicated section. 

Prior to MACRO, excitation/ionization arrays were built on a smaller scale (most of them during the 1980s) at various locations using a variety of different techniques. For these pre-MACRO experiments, see details on \cite{gg84}.
The different searches for bare $\cm$s in the cosmic radiation and their results are listed in Table \ref{tab:GUTtable}.

%%%%%%%%%%%%%%%%%%%%%%%%%%%%%%%%%%%%%%%%%%%%%%%%%%%%%%%%%%%%%%%%%%%%%
\begin{table}
\caption{Flux upper limits for GUT and Intermediate Mass Monopoles from different experiments, assuming $g=g_{D}$.}
\centering 
\resizebox{1.05\textwidth }{!}
{\begin{tabular} {|l|c|c|c|l|}
\hline
Experiment  (Ref) &  Mass Range    & $\beta$ range & Flux Upper Limit                & Detection Technique\\
            & (GeV/c${^2})$  &               & (cm$^{-2}$~s$^{-1}$~sr$^{-1}$)  &                     \\ [0.5ex]
\hline
\hline 
AMANDA II Upgoing \cite{amanda} & $10^{11}-10^{14}$ & $0.76-1$ & $(8.8-0.38)\times 10^{-16}$ & Ice Cherenkov \\
%\hline
AMANDA II Downgoing \cite{amanda} & $10^{8}-10^{14}$ & $0.8-1$ & $(17-2.9)\times 10^{-16}$ & Ice Cherenkov\\
%\hline
IceCube \cite{icecube} & $10^{8}-10^{14}$ & $0.8-1$ & $(5.6-3.4)\times 10^{-18}$ & Ice Cherenkov\\
%\hline
Baikal \cite{baikal} & $10^{7}-10^{14}$ & $0.8-1$ & $(1.83-0.46)\times 10^{-16}$ & Water Cherenkov \\
%\hline
ANTARES \cite{antares} &  $10^{7}-10^{14}$ & $ 0.625-1$ & $(9.1-1.3)\times 10^{-17}$ & Water Cherenkov \\ [0.5ex]
MACRO \cite{02A3} & $5\times10^{8}- 5\times10^{13}$ & $ > 5\times10^{-2} $ & $3\times 10^{-16}$ & Scint.+Stream.+NTDs \\
%\hline
MACRO \cite{02A3} & $>5\times10^{13}$ & $ > 4 \times 10^{-5} $ & $1.4\times 10^{-16}$ & Scint.+Stream.+NTDs \\
%\hline
Soudan 2 \cite{sou2} & $10^8 - 10^{13}$ & $ > 2 \times 10^{-3} $ & $8.7\times 10^{-15}$ & Gas drift tubes \\
%\hline
OHYA \cite{ohya} & $5\times10^{7}- 5\times10^{13}$ & $ > 5\times10^{-2} $ & $6.4\times 10^{-16}$ & Plastic NTDs \\
%\hline
OHYA \cite{ohya} & $>5\times10^{13}$ & $ > 3\times 10^{-2} $ & $3.2\times 10^{-16}$ & Plastic NTDs \\
[0.5ex]
%\hline
SLIM \cite{slim} & $10^{5}- 5\times10^{13}$ & $ > 3\times10^{-2} $ & $1.3\times 10^{-15}$ & Plastic NTDs \\
%\hline
SLIM \cite{slim} & $>5\times10^{13}$ & $ > 4 \times 10^{-5} $ & $0.65\times 10^{-15}$  & Plastic NTDs \\[0.5ex]
%\hline
%\hline 
Induction, combined \cite{gg84} & $>10^{5}$ & any & $4 \times 10^{-13}$ & Induction \\
\hline 
\end{tabular}} 
\label{tab:GUTtable}
\end{table}
%%%%%%%%%%%%%%%%%%%%%%%%%%%%%%%%%%%%%%%%%%%%%%%%%%%%%%%%%%%%%%%%%%%%%

Fig. \ref{fig:under} shows the 90\% CL flux upper limits versus $\beta$ for GUT $\cm$s with $g=g_{D}$ from different searches. 
The experiment at the Ohya mine in Japan \cite{ohya} used a 2000 m$^2$ array of NTDs; 
Baksan in Russia \cite{baksan} used liquid scintillation counters; 
Soudan 2 in USA \cite{sou2} was a large fine-grained tracking calorimeter comprised of long drift tubes, and Kolar Gold Field (KGF) \cite{kgf} was also a tracking calorimeter in India. 
The others experiments are described below.

\subsubsection{Searches at high altitude laboratory}
%%%%%%%%%%%%%%%%%%%%%%%%%%%%%%%%%%%%%%%%%%%%%%%%%%%%%%%%
The SLIM experiment \cite{slim} at the Chacaltaya laboratory (5230 m a.s.l., Bolivia) searched for downgoing IM$\cm$s with a 427 m$^{2}$ NTD array exposed for 4.2 years to the cosmic radiation. SLIM was sensitive to $\cm$s with $g=2g_{D}$ in the whole range $4 \times 10^{-5}<\beta <1$ and $\beta >10^{-3}$ for $g=g_{D}$. 
No candidate event was observed; a 90\% CL flux upper limit of $\sim 1.3\times 10^{-15}$ cm$^{-2}$s$^{-1}$sr$^{-1}$ was set.

\subsubsection{Relativistic magnetic monopoles in neutrino telescopes}
%%%%%%%%%%%%%%%%%%%%%%%%%%%%%%%%%%%%%%%%%%%%%%%%%%%%%%%%

Relativistic charged particles travelling through a homogeneous and transparent medium like water or ice emit Cherenkov radiation that can be detected by arrays or strings of photomultiplier tubes.
Baikal \cite{baikal}, AMANDA \cite{amanda}, ANTARES \cite{antares}, IceCube \cite{icecube}, and in the future KM3NeT are usually denoted as neutrino telescopes \cite{Chiarusi}.
Due to the Cherenkov mechanism, a huge quantity of visible Cherenkov light would be emitted by a $\cm$ with $\beta> c/n= 0.75$, where $n=1.33$ is the refractive index of water \cite{05L1}. Additional light is produced by Cherenkov radiation from $\delta$-ray electrons along the monopole's path for velocities down to $\beta= 0.625$.

A large background from cosmic muons inhibit in most cases the search for down-going candidates.  Up-going monopoles are required to traverse the Earth before reaching the detector. It is worth noting that it is unlike for GUT supermassive $\cm$s to reach (nearly) relativistic velocities.
In the IceCube analysis, the threshold was adjusted for down-going signals according to the direction.

\subsubsection{Ultra-relativistic magnetic monopoles}
%%%%%%%%%%%%%%%%%%%%%%%%%%%%%%%%%%%%%%%%%%%%%%%%%%%%%%%%
Constraints on the flux of ultra-relativistic $\cm$s were also given by two experiments based on the detection of radio wave pulses from the interactions of a primary particle in ice.
The Radio Ice Cherenkov Experiment, RICE, consisting of radio antennas buried in the Antarctic ice set a flux upper limit at the level of $10^{-18}$ cm$^{-2}$~s$^{-1}$~sr$^{-1}$ (95\% CL) for intermediate mass monopoles with a Lorentz factor $10^{7}<\gamma<10^{12}$ and $10^{16}$ GeV  total energy \cite{rice}. 
The ANITA-II balloon-borne radio interferometer determined a 90$\%$ CL flux upper limit of the order of $10^{-19}$ cm$^{-2}$~s$^{-1}$~sr$^{-1}$ for $\gamma>10^{10}$ at the total energy of $10^{16}$ GeV \cite{anita}. 

\subsubsection{Induction techniques}
%%%%%%%%%%%%%%%%%%%%%%%%%%%%%%%%%%%%%%%%%%%%%%%%%%%%%%%%
Induction detectors are considered the most robust method to identify the passage of a magnetic monopole of any velocity, mass and magnetic charge.
In the searches for $\cm$s in cosmic rays, several loops in coincidence are needed to avoid spurious signals. 
The major drawbacks are the cost of the superconductive loops with the required sensitivity, their small effective area (up to a square meter) and the fact that such devices cannot be easily operated in any environment.
A number of searches using apparata with  different number of loops (see for instance \cite{cabrera2,cabrera3}) set a combined flux upper limit of $\sim 4\times 10^{-13}$ cm$^{-2}$sr$^{-1}$s$^{-1}$ (out of the scale of Fig. \ref{fig:under}).
They provide the only direct constraints on the monopole flux for $\beta  < 10^{-4}$.

%%%%%%%%%%%%%%%%%%%%%%%%%%%%%%%%%%%%%%%%%%%%%%%%%%%%%%
\subsection{The MACRO experiment at Gran Sasso}\label{sec:5-macro}
%%%%%%%%%%%%%%%%%%%%%%%%%%%%%%%%%%%%%%%%%%%%%%%%%%%%%%

MACRO (Monopole Astrophysics and Cosmic Ray Observatory) \cite{02A4} was a large multipurpose underground detector located in the Hall B of the Laboratori Nazionali del Gran Sasso (Italy). 
At Gran Sasso the minimum thickness of the rock overburden is 3150 hg/cm$^2$ and the cosmic muon flux is reduced to $\sim 1$ m$^{-2}$ h$^{-1}$, almost a factor 10$^6$ smaller than at the surface.
The detector, which took data with different configurations from 1989 to December 2000, was optimized for the search for GUT magnetic monopoles with velocity $\beta \ge 4\times 10^{-5}$. 
The large area apparatus had global dimensions of $76.6\times 12 \times 9.3$ m$^3$, and $\sim 10,000$ m$^2$sr acceptance to an isotropic particle flux.
Redundancy and complementarity were the primary features in designing the experiment, because reasonably very few $\cm$s during the detector lifetime were expected. 
To accomplish this, the apparatus consisted of three independent sub-detectors: liquid scintillation counters, limited streamer tubes, each of them with dedicated and independent hardware, and nuclear track detectors. 
MACRO was a multipurpose detector, able to address different cosmic ray physics items, as the disappearance of atmospheric $\nu_\mu$, interpreted as due to neutrino oscillations \cite{MAosci}.

Because the signature of the passage of a GUT $\cm$ across the detector depends strongly on its velocity, different hardware systems were designed and applied to maximize the sensitivity at different $\beta$. Multiple analysis strategies were adopted, where both the $\cm$ signatures and the background characteristics were took into account.
Fast $\cm$s were looked for in the active detectors via their large energy releases in the liquid scintillator \cite{MAsci} and in the limited streamer tube \cite{MAst} systems.
The background was mainly due to high-energy muons with or without an accompanying electromagnetic shower. 
Slow magnetic monopoles should leave small signals spread over a large time window \cite{MAall1}; a $\beta \sim 10^{-4}$ monopole would have had a time-of-flight across the detector as large as about 1 ms. 
This implied the use of specific analysis procedures that allowed the rejection of the background mainly due to radioactivity induced hits and possible electronic noise.
The final result was obtained by a proper combination of flux upper limits from different searches, as reported in \cite{02A3}.

\subsubsection{The scintillator system}
%%%%%%%%%%%%%%%%%%%%%%%%%%%%%%%%%%%%%%%%%%%%%%%%%%%%%%%%
%This sub-system was organized in three layers of horizontal and two %layers of vertical counters. 
The response of liquid scintillators (discussed in \S \ref{sec:4-scint}) to heavily ionizing particles was studied experimentally \cite{fice} and their sensitivity to particles with velocity down to $\beta \sim  10^{-4}$ was directly measured.
Slow $\cm$s were searched for using dedicated hardware, the Slow Monopole Trigger and a 200 MHz custom-made Wave Form Digitizer system \cite{MAslo}. 
The system was sensitive over the entire range of pulse widths and amplitudes (which could be just a train of single photoelectron pulses lasting over several microseconds) expected for slow moving $\cm$s while suppressing efficiently narrow (10-50 ns) pulses due to isolated radioactivity and cosmic ray muons.
In addition, the amplitude and time-of-flight information for every particle crossing the counters were recorded by a stand-alone ADC/TDC system and used for fast monopole searches.

\subsubsection{The limited streamer tube system}
%%%%%%%%%%%%%%%%%%%%%%%%%%%%%%%%%%%%%%%%%%%%%%%%%%%%%%%%
The streamer tube system was designed to be effective in the search for $\cm$s in a wide velocity range, $10^{-4} < \beta < 1$. 
For this purpose, a gas mixture containing helium and n-pentane was used to exploit the Drell-Penning effect. 
In the higher velocity region ($\beta > 10^{-3}$), where the assumptions of the Drell-Penning effects do not apply, the standard ionization mechanism expected for a $\cm$ ensures an energy release much larger (see Fig. \ref{fig:lst}b) than that from minimum ionizing particles. 
The charge collected on the central wire of the streamer tube has a logarithmic dependence on the energy released inside the active cell so that a charge measurement allowed one to distinguish between $\cm$s and muons.

\subsubsection{The nuclear track detector system}
%%%%%%%%%%%%%%%%%%%%%%%%%%%%%%%%%%%%%%%%%%%%%%%%%%%%%%%%
The NTD covered a total area of 1263 m$^2$ and was organized in stacks of $24.5\times 24.5$ cm$^2$ consisting each of three layers of CR39, three layers of Lexan and an aluminum absorber placed in an aluminized Mylar bag filled with dry air. 
The CR39 allowed a search for $\cm$s of different magnetic charges and $\beta$. For a single Dirac charge the MACRO CR39 was sensitive in the ranges $2\times 10^{-5}< \beta < 2\times 10^{-4}$ and $\beta > 1.2 \times 10^{-3}$ \cite{derk98}. 
Lexan has a much higher threshold making it sensitive to 
$\cm$s with $\beta > 0.1$.
After about 9.5 y of exposure, the top CR39 foils in each stack were etched and analyzed. 
The signal looked for was a hole or a biconical track with the two base cone areas equal within experimental uncertainties.
%After etching, the foils were scanned twice using back lighting by %different operators at low magnification, looking for any possible %optical inhomogeneity. 
An efficiency close to 100\% for finding a signal was achieved. 
The signature of the passage of a $\cm$ was a coincidence among different CR39 layers. 
Background tracks could be induced by proton recoils from neutron interactions or by low energy nuclei from muon interactions in the material surrounding the apparatus.
No signal was found.

\subsubsection{The results}
%%%%%%%%%%%%%%%%%%%%%%%%%%%%%%%%%%%%%%%%%%%%%%%%%%%%%%%%
No event in MACRO was compatible with the passage of a $\cm$ in either individual sub-systems or in combined analysis \cite{02A1}, apart the ``spokesmen event''. 
That was a LED-induced event (realistically simulating a $\beta \sim 10^{-3}$ $\cm$ in the scintillator system) intentionally and secretly generated to test the efficiency of analyses. 
Both methods designed to detect a $\beta=10^{-3}$ $\cm$ successfully reconstructed the fake ``spokesmen event''. 

The shape of the line labeled MACRO in Fig. \ref{fig:under} reflects the combination of the independent results obtained with the different search methods. 

Fig. \ref{fig:mcata}a) shows the flux upper limits as a function of the mass $M$ for monopole speed $\beta\ge 0.05$ for the MACRO and Ohya underground experiments and for the SLIM high-altitude detector.

%%%%%%%%%%%%%%%%%%%%%%%%%%%%%%%%%%%%%%%%%%%%%%%%%%%%%%
\subsection{Searches for monopoles bound in matter\label{sec:5-bouma}}
%%%%%%%%%%%%%%%%%%%%%%%%%%%%%%%%%%%%%%%%%%%%%%%%%%%%%%
Bulk matter could have trapped incident cosmic monopoles over a long exposure time. Several searches for cosmic $\cm$s bound in bulk matter have been performed.
One technique \cite{price2,ghosh} used few pieces of $\sim 10^9$ old mica samples, chemically etched and scanned in transmitted light under polarizing microscope to search for etch pits originating from elastic collisions of a $\cm$ with a nucleus. The absence of any etch pit was used to set upper limit of $10^{-16}$ to $10^{-18}$ cm$^{-2}$ s$^{-1}$ sr$^{-1}$ on the flux of the supermassive ($M\sim 10^{17}$ GeV) $\cm$s having velocity in the range $3\times 10^{-4} \lesssim \beta\lesssim 1.5\times 10^{-3}$.

Samples of terrestrial, lunar and meteoritic materials were passed through a superconducting magnet looking for a persistent current in the coil induced by the monopole after the passage of the samples.
For instance, a recent search for monopoles in polar igneous rocks \cite{polar} probed 23.4 kg of samples.
% for which a limit on the monopole density of $9.8\times 10^{-5}/$gram %at 90\% CL was set.
Globally, these searches constrain monopole concentrations in matter below 1 per few hundred kg, corresponding to an upper limit of $\cm$ per nucleon ratio of $\sim 10^{-29}$. 
For a complete review, see Sez. 5 of \cite{burdin}.

%%%%%%%%%%%%%%%%%%%%%%%%%%%%%%%%%%%%%%%%%%%%%%%%%%%%%%
\subsection{Searches for monopoles inducing catalysis of nucleon decay}\label{sec:5-cata}
%%%%%%%%%%%%%%%%%%%%%%%%%%%%%%%%%%%%%%%%%%%%%%%%%%%%%%
%% gps

In the hypothesis of a non-negligible cross section, the $\cm$ induced nucleon decay can be exploited to detect the passage of a monopole through series of decay events (f.e. \( \cm + p \rightarrow \cm + e^{+} + \pi^{0}\) via the Rubakov-Callan mechanism) more or less along a straight line.
Early proton decay experiments as Soudan 1 \cite{MCSudan}, IMB \cite{MCimb} and Kamiokande \cite{MCkam} searched for cosmic monopoles and established limits to their flux via the catalysis mechanism.

\begin{figure*}[tb]
%\vspace*{2mm}
\begin{center}
\includegraphics[width=16.0cm]{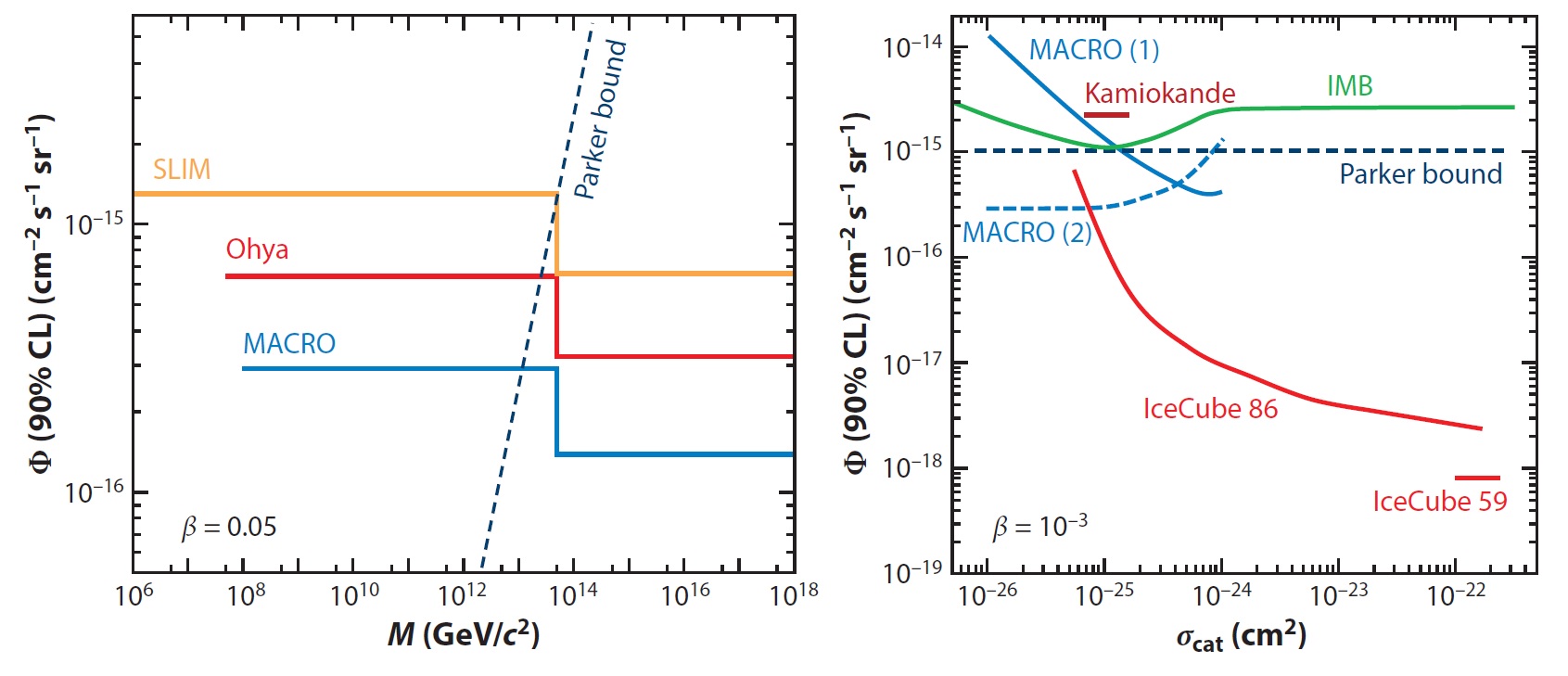}
\end{center}
%\vspace*{-12mm}
\caption{\small\label{fig:mcata} a) Upper flux limits for GUT $\cm$s as a function of their mass $M$ for $\beta=0.05$ as set by MACRO \cite{02A3}, Ohya \cite{ohya} and SLIM \cite{slim}.
b) Upper limits on the flux of $\beta=10^{-3}$ $\cm$ as a function of the catalysis cross section $\sigma_{cat}$ for two IceCube analyses \cite{ic14}, two MACRO analyses \cite{catalisi}, IMB \cite{MCimb} and Kamiokande \cite{MCkam}.}
\end{figure*}

In the MACRO experiment \cite{catalisi}, a dedicated search based on the streamer tube subsystem was carried out.  
A full Monte Carlo simulation of the expected signal was performed.
Different catalysis cross sections were assumed, ranging from $10^{-24}$ cm$^2$ to $10^{-26}$ cm$^2$. 
The null result of the search was used to set limits on the cross section parameters and on the magnetic monopoles flux.
In Fig. \ref{fig:mcata}b) upper limits from different experiments are shown for a reference monopole speed of $\beta=10^{-3}$.

In neutrino telescopes, the Cherenkov light from nucleon decays along the monopole trajectory would produce a characteristic hit pattern.
Early results were obtained by the Baikal experiment \cite{MCbai,Baikal-cata}.
Recently, the IceCube collaboration implemented a dedicated slow-particle trigger in the DeepCore, the central and denser subdetector of IceCube.
The analysis using one year of data \cite{ic14} yields upper limits (90\% CL) for the flux of non-relativistic GUT monopoles at different speeds. 
The limits reported in Fig. \ref{fig:mcata}b) refer to $\beta=10^{-3}$.

%$10^{-18}\ (10^{-17})$ cm$^{-2}$ s$^{-1}$ sr$^{-1}$ at  catalysis cross %sections of $10^{-22}\ (10^{-24})$ cm$^2$.

The Super-Kamiokande \cite{SK} collaboration performed an indirect search for GUT monopoles, looking for a neutrino signal coming from proton decay catalyzed by GUT $\cm$s captured in the Sun. 
The flux upper limit was at $\Phi_M(\sigma_0/1 mb) < 6.3 \times 10^{-24} (\beta/10^{-3})^2 \textrm{ cm}^{-2} \textrm{ s}^{-1} \textrm{ sr}^{-1}$ at 90\% CL, assuming a catalysis cross section $ \sigma_0$ at $\beta=1$. The limit is valid for $M<10^{17}$ GeV/c$^{2}$. 

%%%%%%%%% 
%%%%%%%%% 

%%%%%%%%%%%%%%%%%%%%%%%%%%%%%%%%%%%%%%%%%%%%%%%%%%%%%%%
%%%%%%%%%%%%%%%%%%%%%%%%%%%%%%%%%%%%%%%%%%%%%%%%%%%%%%%
%%%%%%%%%%%%%%%%%%%%%%%%%%%%%%%%%%%%%%%%%%%%%%%%%%%%%%%
\section{Perspectives}
%\vspace{-0.2cm}{\it\small( 2 pages)}
\label{sec:perspective}
%%%%%%%%%%%%%%%%%%%%%%%%%%%%%%%%%%%%%%%%%%%%%%%%%%%%%%%
%%%%%%%%%%%%%%%%%%%%%%%%%%%%%%%%%%%%%%%%%%%%%%%%%%%%%%%
%%%%%%%%%%%%%%%%%%%%%%%%%%%%%%%%%%%%%%%%%%%%%%%%%%%%%%%

The Standard Model (SM) explains extremely well all available experimental results in particle physics.
The SM predictions had precise confirmations from the measurements performed at every accelerator or collider of increasing energy, culminating with the discovery at the CERN LHC of the last missing piece of the theory: the Higgs boson.
On the other hand, most physicists believe that the SM is incomplete and represents a sort of low energy limit of a more fundamental theory, which should reveal itself at higher energies.

The threshold for this higher energy limit could be so high that no accelerator on Earth, also in the far future, would be able to reach it. 
It is in this context that the searches for magnetic monopoles play a fundamental role.
Such experiments, as well as those searching for proton decay, motivated the birth of large underground experiments in the 1980s contributing significantly to the development of astroparticle physics \cite{spurio}. 
This field is giving a Pandora of new and unexpected results, exploiting the experimental techniques also used at accelerators in the study of the cosmic radiation, in neutrino physics and astrophysics, in the searches for rare phenomena.

%\paragraph LHC}
%%%%%%%%%%%%%%%%%%%%%%%%%%%%%%%%%%%%%%%%%%%%%%%%%%%%%%%
In 2015, following a long technical shutdown, LHC will start to collect data taking. The LHC experiments, all equipped with exceptional detectors, should be able to set new stringent limits on the production cross sections for magnetic monopoles.
Three complementary techniques have been proposed to search for highly ionizing particles, including classical monopoles in a wide range of magnetic charges: in-flight detection in general-purpose apparata; in-flight detection using NTDs; the detection of stopped monopoles in matter with the induction technique \cite{roeck}. 

The in-flight signature of a monopole event in ATLAS \cite{pr-atlas} and CMS \cite{pr-cms} would be rather striking, with a large number of high-threshold hits in the tracker and a very localized energy deposition in the electromagnetic calorimeter \cite{lhc+cos}. 
New upper limits on the production of exotic particles with masses in the TeV energy range can be set.

MoEDAL \cite{moedal} aims to $\cm$'s in-flight detection using NTD stacks deployed around the Point-8 intersection region of the LHCb detector. 
The array will cover a surface of $\sim 25$ m$^{2}$. Each stack, $25 \times 25$ cm$^{2}$, will consist of 9 interleaved layers of CR39, Makrofol and Lexan NTDs.
The $\cm$ signature in MoEDAL is a sequence of collinear etch-pits consistent with the passage of a particle with constant energy loss through all the detector foils. 
The experiment is expected to be sensitive to monopole production down to cross sections of the order of 1 fb.
The sensitivity and redundancy of MoEDAL is enhanced by the addition of a Magnetic Monopole Trapper consisting of an aluminium array: monopoles stopping within this array could be revealed by the induction technique.

Finally, another indirect search is planned through the analysis of exposed beam pipe material using SQUIDs.
The observability of monopoles and monopolium, the monopole-antimonopole bound state \cite{83H3}, in the $\gamma \gamma$ channel for monopole masses in the range 500-1000 GeV is considered in \cite{epele}.

%%%%%%%%%%%%%%%%%%%%%%%%%%%%%%%%%%%%%%%%%%%%%%%%%%%%%%%%%%%%%%%%

The interconnection among cosmic rays studies, neutrino physics and the search for large mass $\cm$s is evident also in future planned experiments, where large area neutrino telescopes and long baseline neutrino experiments can extend the observation in the space of parameters ($M,\beta$) in which monopoles can be sought for. 

%% Nova
%%%%%%%%%%%%%%%%%%%%%%%%
The NO$\nu$A detector is designed to study $\nu_\mu \rightarrow \nu_e$ oscillations in the existing Fermilab NuMI neutrino beam. The far detector, with a volume of $15.5\times 15.5 \times 59.5$ m$^3$, is located near the surface in northern Minnesota, USA. 
When completed, it will contain $\sim 3.4\ 10^5$ cells filled with liquid scintillator read out at both ends by avalanche photodiodes.
The far detector, due to its surface proximity, large size, good timing resolution, large energy dynamic range, and continuous readout, is sensitive to the passage of cosmic $\cm$s over a large range of velocities ($10^{-4} <\beta <1$) and masses. 
Two dedicated software-based triggers have been designed \cite{nova} in order to record fast and slow $\cm$ events with high efficiency and a large rejection factor against a rate of cosmic rays of about $\sim 10^5$ Hz.

%% INO
The Iron CALorimeter (ICAL) at India-based Neutrino Observatory (INO) is another experiment for a precise measurement of the neutrino oscillation parameters.
Magnetic monopoles will be searched for in ICAL \cite{ino} using Resistive Plate Chambers (RPCs).
The RPCs yield saturated pulses carrying the hit time information, allowing to track particles crossing the apparatus. 
The Drell-Penning effect might not be effective in the ICAL gas system, therefore the experiment would be sensitive to magnetic monopoles with $\beta >10^{-3}$. 
The highest allowed velocity is $\beta \sim 0.9$, limited by the huge background of atmospheric muons.
The expected sensitivities of NO$\nu$A and ICAL are however not significantly larger than that reached by MACRO.

In order to improve significantly upon the existing upper limits on the $\cm$ flux, a next generation of experiments covering very large surfaces are required, possibly order of 50-100 times larger acceptance than MACRO (and SLIM), based on simple and low cost detector designs. 
Achieving this improvement will not be a simple task.

The very large ionization energy losses of $\cm$s in a wide $\beta$ range suggest detectors with high $dE/dx$ thresholds to ease the discrimination of the backgrounds. 
NTDs are relatively cheap detectors that can cover very large surfaces. The disadvantage of NTDs is that they require chemical etching and optical scanning to extract the signal. 
Solid state breakdown counters, that exhibit high registration $dE/dx$ thresholds, may offer an alternative to NTDs, providing in addition the convenience of electronic registration and simplicity of fabrication, operation, and signal extraction. 
These properties make such technique a viable one to search for $\cm$s and other highly ionizing rare particles \cite{solid}.

Neutrino telescopes are the largest detectors (area $\sim 10^6$ m$^2$) that could record a long sequence of hits along a particle trajectory. 
They can detect relativistic magnetic monopoles, emitting radiation above the Cherenkov threshold; alternatively, they can catch the light sequence of a monopole catalyzing proton decay processes.  
These detectors can reach a sensitivity three order of magnitudes below the Parker bound, but in a very limited interval of $\beta$ or under the assumption of the Rubakov-Callan mechanism.

%%% Macrosistemi
In recent years a number of phenomena in condensed matter have been observed as manifestations of the correlations present in a strongly interacting many-body systems, behaving like magnetic monopoles (e.g. \cite{qp1,qp2}).  These systems are called {}``emergent particles'' or {}``quasiparticles''. 
In January 2014 the synthesis of a Dirac monopole quasiparticle in a spinor Bose-Einstein condensate was reported. 
It was obtained by using an external magnetic field to guide the spins of the atoms forming the condensate \cite{qp3}. 
%Such systems are not fundamental particles, and they do not bear on %electric charge quantization or Grand Unified Theories. 
The attention aroused by such findings as well as the many searches performed, those planned, and the abundant theoretical literature indicate the persistent lively interest on the possible existence of the magnetic monopole. As Dirac, still many people today ``would be surprised if Nature had made no use of it'' (\cite{dirac}, pag. 71).

%%%%%%%%%%%%%%%%%%%%%%%%%%%%%%%%%%%%%%%%%%%%%%%%%%%%%%%%%%%%%
%%% ACKNOWLEDGMENTS and References
%%%%%%%%%%%%%%%%%%%%%%%%%%%%%%%%%%%%%%%%%%%%%%%%%%%%%%%%%%%%%

\section*{ACKNOWLEDGMENTS}
We are indebted  to dr. Z. Sahnoun, for her precious help in the preparation of this review.
We thank our colleagues of the MACRO and SLIM collaborations.
We acknowledge the contributions of many colleagues in different searches for magnetic monopoles in the last 25 years. 
In particular S. Balestra, S. Cecchini, M. Cozzi, M. Giorgini, A. Kumar, G. Mandrioli, S. Manzoor, A. Margiotta, V. Popa, M. Sioli, G. Sirri and V. Togo. 
The leader of most of those adventures was Giorgio Giacomelli, former MACRO co-spokesperson, passed away on January 30th, 2014.
To our mentor and deeply missed colleague, this review, largely based on his lessons and reviews, is dedicated.

%%%%%%%%%%%%%%%%%%%%%%%%%%%%%%%%%%%%%%%%%%%%%%%%%%%%%%%%%%%%%

\end{document}